\documentclass[12pt]{article}

\newenvironment{wileykeywords}{\textsf{Keywords:}\hspace{\stretch{1}}}{\hspace{\stretch{1}}\rule{1ex}{1ex}}

\usepackage{amsmath,amssymb}
\usepackage{graphicx}
\usepackage{color}
\usepackage{dcolumn}
\usepackage{bm}
\usepackage[numbers,super,comma,sort&compress]{natbib}

\definecolor{background-color}{gray}{0.98}

\title{Calculating particle pair potentials from fluid-state pair correlations:\\Iterative Ornstein--Zernike Inversion}
\author{Marco Heinen\thanks{mheinen@fisica.ugto.mx
Departamento de Ingenier\'{i}a F\'{i}sica,
Divisi\'{o}n de Ciencias e Ingenier\'{i}as,
University of Guanajuato,
Loma del Bosque 103,
37150 Le\'{o}n,
M\'{e}xico}}

\begin{document}

\maketitle

\begin{abstract}
An iterative Monte Carlo inversion method for the calculation of particle pair potentials from given particle pair
correlations is proposed in this paper. The new method, which is best referred to as \emph{Iterative Ornstein-Zernike Inversion},
represents a generalization and an improvement of the established \emph{Iterative Boltzmann Inversion} technique
[Reith, P\"{u}tz \& M\"{u}ller-Plathe, \emph{J. Comput. Chem.} \textbf{24}, 1624 (2003)]. Our modification of Iterative Boltzmann Inversion
consists of replacing the potential of mean force as an approximant for the pair potential with another, generally more accurate approximant
that is based on a trial bridge function in the Ornstein-Zernike integral equation formalism. As an input, the new method requires the particle
pair correlations both in real space and in the Fourier conjugate wavenumber space.
An accelerated iteration method is included in the discussion, by which the required number of iterations can be greatly reduced
below that of the simple Picard iteration that underlies most common implementations of Iterative Boltzmann Inversion.  
Comprehensive tests with various pair potentials
show that the new method generally surpasses the Iterative Boltzmann Inversion method in terms of reliability of the
numerical solution for the particle pair potential.
\end{abstract}

\begin{wileykeywords}
Monte Carlo Simulation, Ornstein-Zernike Equation, Effective Interactions, Coarse Graining, Iterative Boltzmann Inversion.
\end{wileykeywords}

\clearpage


\section*{\sffamily \Large INTRODUCTION} 

Henderson's theorem \cite{Henderson1974} states that
\emph{'the pair potential [$v(r)$] which gives rise to a given radial distribution function $g(r)$ is unique up to a constant'}
for equilibrium fluids in which the total potential energy is the sum of the interaction energies in all pairs of particles.
It is therefore possible in principle to determine the particle pair potential, denoted $u(r)$ in the present paper,
if only the function $g(r)$ is known.
Nonetheless, great problems are encountered in the practical application of algorithms that aim to deduce $u(r)$ from a given target
function $g_T(r)$ that exhibits statistical or systematic uncertainty \cite{Potestio2014}:   
The functional mapping $u(r) \to g(r)$ is highly nonlinear, causing a propagation and amplification of small errors in $g_T(r)$ into large errors of $u(r)$
if an inverse mapping $g_T(r) \to u(r)$ is attempted. The resulting potential $u(r)$ may reproduce the input function $g_T(r)$ perfectly within its
uncertainty level, but this does not imply that $u(r)$ is close to the true particle pair potential in the system from which $g_T(r)$
was originally obtained. The core problem is that very different pair potentials $u(r)$ can result in very similar functions $g(r)$, all of which agree well
with $g_T(r)$.
In some applications, such as coarse graining of a multicomponent simulation, it may be acceptable to obtain just one of the different potentials
that reproduce the pair correlations of one of the species. In other situations, such as the analysis of experimentally recorded pair correlations,
one is often interested in the precise form of the true interaction potential of the observed particles. In such cases, a reliable method is required
that is capable of picking the true potential from a family of structure-reproducing candidates $u(r)$.
This picking of the optimal candidate is a non-trivial task that requires careful algorithm design and optimal use of the available information.

\begin{figure}
\centering
\includegraphics[width=.4\columnwidth]{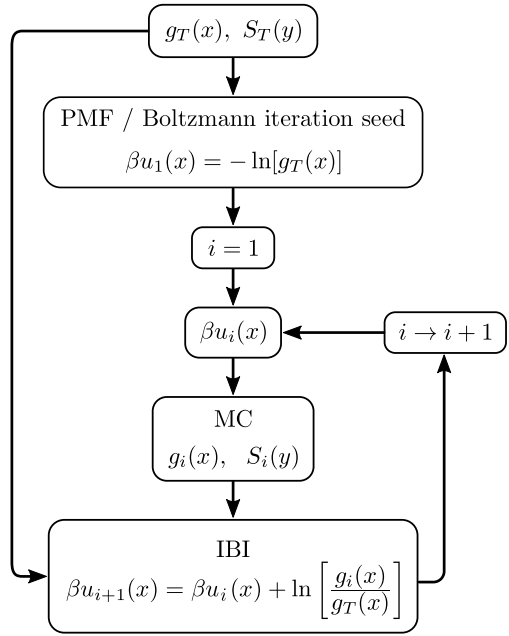}
\vspace{0em}
\caption{\label{fig:IBI_Flowchart}
Flowchart representation of
Iterative Boltzmann Inversion (IBI)\cite{Reith2003}
in the most simple and common form of a Picard iteration.
The method uses a radial distribution function $g_T(x = r n^{1/3})$ as input for iterating
an approximation $\beta u_i(x)$ of the reduced, dimensionless particle pair potential.
Both the iteration seed $\beta u_1(x)$ and the iteration update rule (in the lowermost box)
represent $\mathcal{O}(n)$ approximations by neglecting the difference between the potential
of mean force (PMF) $-k_B T \ln[g_T(x)]$ and the true pair potential $u(x)$.
The complementary information contained in the static structure factors $S_T(y = q / n^{1/3})$ and $S_i(y)$ 
is not used in the iteration.
}
\end{figure}

One prominent algorithm for the calculation of $u(r)$ from a given $g_T(r)$ is Iterative Boltzmann Inversion (IBI)\cite{Reith2003}, a flowchart
representation of which is provided in fig.~\ref{fig:IBI_Flowchart}. 
The working principle of IBI is an initial approximation of the true particle pair potential $u(r)$ by the potential of mean force (PMF)
$w(r) = -k_B T \ln[g_{T}(r)]$, followed by an iterative improvement of $u(r)$. Here, $k_B$ is the Boltzmann constant and $T$ denotes the absolute temperature.
Using the symbol $\beta = 1 / (k_B T)$ for the inverse thermal energy, the IBI algorithm commences with the calculation of the iteration seed
$\beta u_1(r) = \beta w(r)$ for a given function $g(r) = g_T(r)$, as represented by the second
box from the top in fig.~\ref{fig:IBI_Flowchart}. Note that we use the symbols $x = r n^{1/3}$ and $y = q n^{-1/3}$
for the reduced, dimensionless particle center-to-center distance $x$ and its Fourier conjugate variable, the reduced dimensionless wavenumber $y$
in the rest of this paper. These dimensionless variables result from using the mean geometric particle distance $n^{-1/3}$ as a unit of length,
where $n = N/V$ is the number density of $N$ particles in a $3$-dimensional system of volume $V$.

The iteration seed $\beta u_1(x)$ is used in a Monte Carlo (MC) simulation from which a radial distribution function $g_1(x)$ is extracted.
Using $i = 1, 2, 3, \ldots$ as an iteration index and approximating $\beta u_{i+1}(x) \approx -\ln[g_T(x)]$ and $\beta u_{i}(x) \approx -\ln[g_{i}(x)]$
results in the IBI update rule
\begin{equation}\label{eq:IBI_update}
\beta u_{i+1}(x) = \beta u_{i}(x) + \ln\left[\dfrac{g_i(x)}{g_T(x)} \right],
\end{equation}
as contained in the lowermost box in fig.~\ref{fig:IBI_Flowchart}.
Running a new MC simulation in every iteration, the reduced, dimensionless potential $\beta u_i(x)$ is iterated until convergence,
at which point it represents the output function of the IBI algorithm.
For the reasons mentioned above, IBI is prone to result
in a potential that is distinctly different from the true potential $u(x)$ in the system from which $g_T(x)$ has been obtained.
Moreover, it has been shown \cite{Jain2006} that IBI is a creeping process which can take hundreds of iterations before a 
converged solution is obtained. With every iteration involving a separate MC simulation, this slow convergence of IBI can give
rise to unacceptably long algorithm runtimes.
This weakness of IBI, demonstrated in the present paper by our results for various test potentials,
calls for a modification of the iteration procedure that improves its speed and its reliability to converge close to the true pair potential.

One useful improvement of standard IBI is the Multistate IBI method\cite{Moore2014} which uses a set of more than one function $g_T(x)$ as its input,
each of which has been obtained at a different thermodynamic state point of the fluid. Multistate IBI determines a potential that reproduces all of the
input functions $g_T(x)$ at their respective state points and thereby greatly reduces the likelihood for misrepresentation of the true potential.
By construction, Multistate IBI assumes that the true pair potential $u(x)$ does not depend on the thermodynamic state of the fluid.
This assumption is practically perfectly valid for a large class of molecular fluids, but it can be violated in general.
In particular, the effective pair potential of mesoscopic, Brownian particulates in coarse-grained descriptions typically does
depend on the thermodynamic state of the solvent \cite{Dijkstra2000, Louis2001, Dobnikar2006, Trizac2007, HeinenPalbergLoewen2014},
rendering Multistate IBI generally inapplicable for such systems.
The remaining need for improvement of IBI in cases where the potential is state-dependent is addressed in the present work.

Another way to improve the precision of IBI consists in a 'pressure correction', which has also been referred to as 'ramp correction'
\cite{Reith2003, Jain2006, Rosenberger2016}. In pressure-corrected IBI, the iterated potential is modified by addition of a linear ramp function
that depends on a single scalar parameter (the slope of the ramp). This parameter is adjusted numerically until the (virial) pressure
of the fluid with the iterated potential matches that of the target fluid. As a result, the converged iterated potential tends to be in
better general agreement with the true potential of the target fluid than the converged potential from non-pressure-corrected IBI.
The pressure correction can be generalized by adding in place of a linear ramp a higher order polynomial to the potential, and fixing each of the polynomial
coefficients by matching a separate thermodynamic observable such as the pressure, the internal energy, several Kirkwood-Buff integrals \textit{et cetera}
\cite{Jain2006, Rosenberger2016}. While the pressure correction is a powerful tool for the improvement of IBI, it is not without problems:
First, it requires information on thermodynamic properties of the target fluid as an input, in addition to the function $g_{T}(x)$. Such thermodynamic
information may not be available in all cases. Second, the ramp (or polynomial) correction represents a rather ad-hoc modification of the
iterated potential that lacks a fundamental physical motivation.

We present here a modification of IBI that uses pair-correlation function input for a single thermodynamic state point only,
based on the key idea of exploiting a target static structure factor $S_T(y)$ \cite{Hansen_McDonald1986} as a complementary, Fourier-space source of 
information together with the real space information from $g_T(x)$. The necessary link between the Fourier and real space
functions is provided by the Ornstein--Zernike (O--Z) equation in conjunction with a closure relation for the generally unknown bridge function.
The new algorithm is therefore must accurately described as Iterative Ornstein--Zernike Inversion (IO--ZI) or
as Iterative Hypernetted Chain Inversion (IHNCI), in the special case in which the Hypernetted Chain\cite{Morita1958} (HNC) closure relation is used.
Making an approximation of $\mathcal{O}(n^2)$ at the level of the bridge function, the IO--ZI method is generally surpassing the IBI method in terms of precision,
since the central approximation $u(r) \approx w(r)$ in IBI causes a larger error of $\mathcal{O}(n)$\cite{Hansen_McDonald1986}.
Without undertaking the pertinent effort here, we note that IO--ZI could be straightforwardly generalized to include the concepts
of Multistate inversion\cite{Moore2014} or pressure correction \cite{Reith2003, Jain2006, Rosenberger2016},
thereby further enhancing the reliability of the method in cases where the potential is state-independent or
where thermodynamic information of the target fluid is available.

In the work presented here, IBI and IO--ZI have been implemented both in their simplest Picard iteration form and in form of a faster convergent fixed-point
iteration scheme which constitutes a generalization of a method that has been introduced by Ng \cite{Ng1974} (see also \cite{Pulay1980, Pulay1982, Heinen2014}).
The resulting 'Ng-accelerated' versions of IBI and IO--ZI are converging to particle pair potentials which are close, but not identical to the 
fixed point solutions of the corresponding Picard iterations.

This paper is organized as follows:
In the methodology section we present the IO--ZI algorithm, focusing on the IHNCI special case. Both the simple Picard iteration
version and the Ng-accelerated version of the algorithm are discussed.
The results section features our comprehensive test cases in which the performance of IHNCI is compared to that of IBI.
The paper concludes with a discussion of the principal advantages and disadvantages of both IBI and IHNCI, including a recommendation
on which method should be used under given circumstances.

\section*{\sffamily \Large METHODOLOGY}
Let $r = |\mathbf{r}|$ denote the norm of the vector $\mathbf{r}$ that connects two particle center points in $3$-dimensional space.
Then, the wavenumber $q = |\mathbf{q}|$ is the conjugate variable to $r$ in the Fourier transform pair
\begin{eqnarray}
\tilde{f}(q) = &\mathcal{F}[f(r)](q)&              =~ ~~\dfrac{4 \pi}{q}~   \int\limits_0^{\infty} ~ dr ~ r ~ f(r) \sin(qr),        \label{eq:Fouriertrans_q_to_r}\\
        f(r) = &\mathcal{F}^{-1}[\tilde{f}(q)](r)& =~ \dfrac{1}{2 \pi^2 r} \int\limits_0^{\infty} ~ dq ~ q ~ \tilde{f}(q) \sin(qr), \label{eq:Fouriertrans_r_to_q}
\end{eqnarray}
for an isotropic function $f(\mathbf{r}) = f(r)$, $\tilde{f}(\mathbf{q}) = \tilde{f}(q)$.
For simplicity we assume isotropy in all functions, restricting thereby the applicability of the presented methods to systems
in which both the particle pair potential $u(r)$ and the particle pair correlation functions $g(r)$, $S(q)$ are isotropic.
Yet, this isotropy assumption is merely a technical simplification that could be lifted in future generalizations of the present work by the
use of technically more involved anisotropic Ornstein-Zernike formalisms \cite{Fries1985, Plischke1986, Kjellander1991, Nygard2012, Chavez-Paez2003, Contreras-Aburto2010, Jaiswal2014}.
These more advanced numerical techniques are expected to extend the applicability range of the present method to systems with anisotropic
particle pair interactions \cite{Fries1985} as well as systems with anisotropic particle correlations resulting from external fields or confinement
\cite{Fries1985, Plischke1986, Kjellander1991, Nygard2012, Chavez-Paez2003, Contreras-Aburto2010} and crystalline systems \cite{Jaiswal2014}.
Using the O--Z formalism implies that the presented method is strictly applicable only to systems in thermodynamic equilibrium \cite{MendozaMendez2017}.
Nevertheless, we note that O--Z equation solutions for isotropic undercooled liquids are routinely and successfully used as input for theories
of slow dynamics and vitrification such as mode-coupling theory \cite{Goetze2009} and self-consistent generalized Langevin equation theory \cite{MendozaMendez2017}.
This indicates that the procedure presented here might also be of use in certain out-of-equilibrium cases such as undercooled liquids, gels or glasses.

The O--Z equation in its common form \cite{Hansen_McDonald1986, Nagele1996}
\begin{equation}\label{eq:O-Zstandard}
g(x) - 1 = c(x) + \int d^3 x' ~ c(x') \left[ g(|\mathbf{x} - \mathbf{x'}|) - 1 \right]
\end{equation}
relates the direct correlation function $c(x)$ to the radial distribution function $g(x)$.
The latter can be defined as
\begin{equation}\label{eq:gx}
g(x) = \dfrac{1}{N} \left\langle \sum\limits_{\substack{i,j = 1\\i \neq j}}^{N} \delta(\mathbf{x} - \mathbf{x}_i + \mathbf{x}_j) \right\rangle
\end{equation}
in terms of the dimensionless particle center coordinates $\mathbf{x}_i = \mathbf{r}_i n^{1/3}$, the Dirac delta function $\delta$,
and the equilibrium ensemble average $\left\langle \ldots \right\rangle$.
It is useful for our purposes to re-write the O--Z equation in the form \cite{Heinen2011, Heinen2011err}
\begin{equation}\label{eq:O-Znonstandard}
c(x) = g(x) - 1 - \mathcal{F}^{-1} \left\lbrace \dfrac{ {\left[S(y) - 1\right]}^2 }{S(y)} \right\rbrace (x)
\end{equation}
which is equivalent to eq.~\eqref{eq:O-Zstandard} and in which $S(y) = 1 + \mathcal{F}[g(x) - 1](y)$ is the static structure factor.
Note that the Fourier integrand ${\left[S(y) - 1\right]}^2 / S(y)$ in eq.~\eqref{eq:O-Znonstandard} is a quickly decaying function
of $y$, which is beneficial in all practical applications in which $S(y)$ is typically only known for a restricted range of $y$ values. 
 
Equation~\eqref{eq:O-Znonstandard} is exact, at the cost of involving the generally unknown function $c(x)$.
Function $c(x)$ can be decomposed as the sum \cite{Hansen_McDonald1986}
\begin{equation}\label{eq:Directcorr_and_Bridgefunc}
c(x) = -\beta u(x) + g(x) - 1 - \ln\left[ g(x) \right] + b(x)
\end{equation}
in which $b(x)$ is the bridge function.
In default of an exact, closed form expression for $b(x)$, the bridge function is commonly approximated by a closure relation \cite{Nagele1996}.
The most typically used O--Z closures are semi-empirical approximations of $b(x)$, often times based on an Ansatz that 
is exact in certain limiting or special cases such as low density, high temperature or one-dimensional systems.
Relying on the approximate validity of a closure relation, eqs.~\eqref{eq:O-Zstandard}, \eqref{eq:Directcorr_and_Bridgefunc}
can be solved numerically \cite{Heinen2014} with $\beta u(x)$ as input, resulting in approximate solutions for $g(x)$ and $S(y)$.
This approach, in which $\beta u(x)$ is considered as a known input, constitutes the standard \emph{forward direction} of solving the
O--Z equation.

In the present work, we employ the O--Z equation in the opposite \emph{backward direction} by treating $g(x) = g_T(x)$ and $S(y) = S_T(y)$ as an input
and solving eqs.~\eqref{eq:O-Znonstandard}, \eqref{eq:Directcorr_and_Bridgefunc} for $\beta u(x)$, under the assumption that $b(x)$ is well approximated
by a certain closure relation. In a practical application, the target correlation functions $g_T(x)$ and $S_T(y)$ can be measured in an experiment,
such as confocal microscopy of a colloidal suspension \cite{Prasad2007}, or they can refer to the correlations among one species of particles in a 
computer simulation of a multicomponent system \cite{Dijkstra2000}. In both cases, inverting the O--Z equation with a closure relation results in an approximation $\beta u(x)$ for the effective
(or coarse grained) particle pair potential, which quantifies the dimensionless interaction energy for two particles of the same species,
after the degrees of freedom of all other particle species have been integrated out. 

Throughout this work, we use the simple hypernetted chain\cite{Morita1958} (HNC) closure relation \mbox{$b(x) \equiv 0$}.
We have tested alternative closure relations (including the Percus-Yevick \cite{Percus1958} and Kinoshita \cite{Kinoshita2003} closures)
without observing a systematic advantage over HNC for any of them, and refrain from a presentation of the pertinent results for the sake of brevity.
With $b(x) = \mathcal{O}(n^2)$, the HNC closure represents a second-order approximation in the density \cite{Hansen_McDonald1986}
that should be expected to exceed the
precision
of the $\mathcal{O}(n)$ approximation $g(x) \approx \exp \left\lbrace -\beta u(x) \right\rbrace$     
which is the foundation of the IBI method. 
\subsection*{\sffamily \large Iterative HNC Inversion}\label{subsec:IHNCI}
\begin{figure}
\centering
\includegraphics[width=.425\columnwidth]{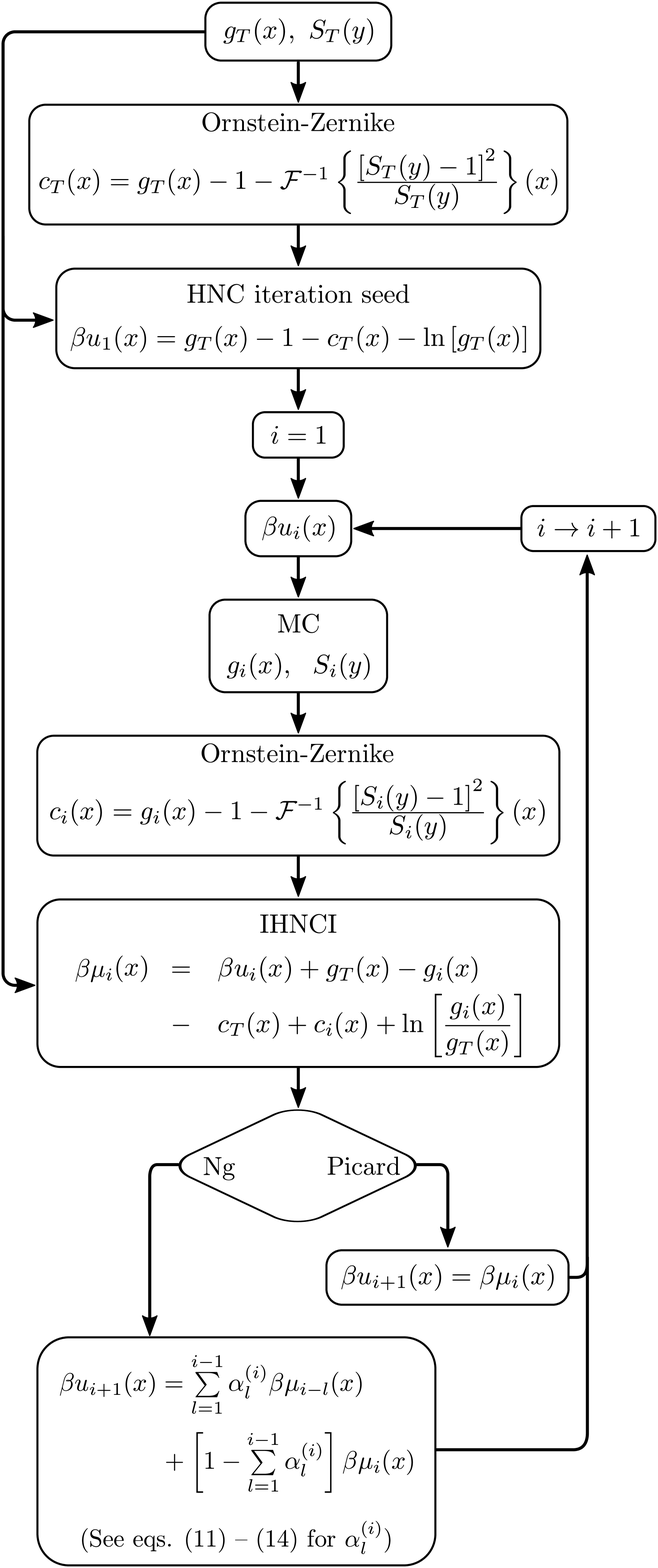}
\vspace{0em}
\caption{\label{fig:IHNCI_Flowchart}
Flowchart representation of the Iterative Ornstein-Zernike Inversion method, which is displayed here in its
Iterative Hypernetted Chain Inversion (IHNCI) form. The method uses a radial distribution
function $g_T(x = r n^{1/3})$ and a static structure factor $S_T(y = q / n^{1/3})$ as input for iterating
an approximation $\beta u_i(x)$ of the reduced, dimensionless particle pair potential.
Both the iteration seed $\beta u_1(x)$ and the iteration update rule (in the box labeled 'IHNCI')
are based on the $\mathcal{O}(n^2)$ Hypernetted Chain approximation which neglects a non-zero bridge function.
}
\end{figure}
Figure~\ref{fig:IHNCI_Flowchart} features a flowchart of the Iterative Ornstein-Zernike Inversion (IO--ZI) algorithm, using the HNC closure as the central approximation.
This special case of IO--ZI is most precisely described as Iterative HNC Inversion (IHNCI), and proceeds as follows:
A target radial distribution function $g_T(x)$ and a target static structure factor $S_T(y)$ of a homogeneous and isotropic equilibrium fluid system are required as the only inputs.
From $g_T(x)$ and $S_T(y)$, the target direct correlation function $c_T(x)$ is calculated via solution of the O--Z eq.~\eqref{eq:O-Znonstandard}, by means
of numerical inverse Fourier transformation.
\subsubsection*{\sffamily \large Numerical inverse Fourier transform}\label{subsubsec:Num_Inv_Fourier}
Using Hamilton's fast and precise FFTLog algorithm \cite{Hamilton2000, Hamilton_website}, based on the original work of Talman \cite{Talman1978},
for the inverse numerical Fourier transform on logarithmically spaced grids in $x$- and $y$-space has the advantage that IO--ZI can be easily
modified for its application to systems in arbitrary spatial dimensions \cite{Heinen2014, Heinen2015}, including non-integer fractal dimensions \cite{Heinen2015PRL}.
In all cases presented here, we have chosen grids with $8192$ points in the intervals $10^{-6} \leq x \leq 10^{6}$ and $4.99 \times 10^{-6} \lesssim y \lesssim 4.99 \times 10^{6}$. 
Since the functions $g_T(x)$ and $S_T(y)$ are usually not given on logarithmic grids, they are interpolated onto the grids by weighted linear
least squares regression of second degree polynomials. The data point weights are sliding Gaussian functions, centered at the
respective grid point to which the input function is being interpolated. After empirical optimization, Gaussian functions in the dimensionless distance ($x$)
and wavenumber ($y$) spaces have been chosen with full widths at half maximum $\Delta x = 0.017$ and $\Delta y = 1.25$. This procedure results in
interpolated functions $g_T(x)$ and $S_T(y)$ that are sufficiently smooth to avoid spurious artifacts in numerical Fourier transformation, yet sufficiently
accurate in their representation of all relevant input function features, since the interpolation is not 'leveling out' the function features by an overly aggressive data smoothing.
Nonetheless, the interpolation procedure is prone to an unphysical smoothing of discontinuities in $g_T(x)$ which occur in case of discontinuous
potentials $u(x)$. We therefore restrict our attention in the present work to particles with continuous potentials and continuous correlation functions.   
\subsubsection*{\sffamily \large Iteration procedure}\label{subsubsec:IHNCI_loop}
After $c_T(x)$ has been determined, the IHNCI algorithm continues with the calculation of the iteration seed
\begin{equation}\label{eq:IHNCI_seed}
\beta u_1(x) = g_T(x) - 1 - c_T(x) - \ln\left[ g_T(x) \right]
\end{equation}
(third box from the top in fig.~\ref{fig:IHNCI_Flowchart}), which represents nothing else than the HNC closure.
An iteration with loop counter $i = 1, 2, 3, \ldots$ is then started, in which the reduced pair potential $\beta u_i(x)$  
is the input for a new (N,V,T) Metropolis MC simulation of a single species fluid in every round.
From each MC simulation, the pair correlation functions $g_i(x)$ and $S_i(y)$ are extracted, which are then used to calculate
$c_i(x)$ via the O--Z eq.~\eqref{eq:O-Znonstandard} and inverse numerical Fourier transformation, as described above for $c_T(x)$.

An IHNCI output potential $\mu_i(x)$ is calculated according to
\begin{equation}\label{eq:IHNCI_update}
\beta \mu_{i}(x) = \beta u_{i}(x) + g_T(x) - g_i(x) - c_T(x) + c_i(x) + \ln\left[\dfrac{g_i(x)}{g_T(x)} \right]
\end{equation}
and, if the simple Picard iteration type of the algorithm is chosen, the updated potential is calculated as $u_{i+1}(x) = \mu_{i}(x)$.

In place of Picard iteration one may alternatively choose the Ng-accelerated iteration type, where the input potential $u_{i+1}(x)$
for a new turn is calculated as the weighted sum
\begin{equation}\label{eq:Ng_iteration}
\beta u_{i+1}(x) =
\sum\limits_{l=1}^{i-1} \alpha_l^{(i)} \beta \mu_{i-l}(x) + 
\left[ 1 -\sum\limits_{l=1}^{i-1} \alpha_l^{(i)}\right] \beta \mu_{i}(x) 
\end{equation}
of all previous IHNCI outputs.
In eq.~\eqref{eq:Ng_iteration}, the scalar mixing coefficients $\alpha_l^{(i)}$ are the entries of the $(i-1)$--component vector
$\underline{\alpha}^{(i)} = (\alpha_1^{(i)}, \alpha_2^{(i)}, \ldots \alpha_{i-1}^{(i)})$ that solves
the linear set of equations
\begin{equation}\label{eq:Ng_linsys}
\underline{\underline{M}}^{(i)} \cdot \underline{\alpha}^{(i)} = \underline{\delta}^{(i)}
\end{equation}
with the symmetric square coefficient matrix $\underline{\underline{M}}^{(i)}$ whose elements are the scalar products
$M^{(i)}_{xy} = \left( d_x^{(i)}, d_y^{(i)} \right)$. Likewise, the right hand side vector $\underline{\delta}^{(i)}$ of
the equation system~\eqref{eq:Ng_linsys} consists of the elements 
$\delta^{(i)}_{x} = \left( d_i, d_x^{(i)} \right)$.
We are using the definitions
\begin{eqnarray}
\left( f, g \right) &=& \int\limits_{x_{\text{min}}}^{x_{\text{max}}} dx ~ f(x) g(x), \label{eq:scalarprod}\\
d_{i}(x) &=& \beta\mu_{i}(x) - \beta u_{i}(x) \label{eq:Ng_d_i_functions}
\end{eqnarray}
and
\begin{equation}
d_{l}^{(i)}(x) = d_{i}(x) - d_{i-l}(x) \label{eq:Ng_d_l^i_functions}  
\end{equation}
for the scalar product of two real-valued functions $f(x)$ and $g(x)$
and for the difference functions $d_{i}(x)$ and $d_{l}^{(i)}(x)$.
A vanishing function $d_{i}(x) \equiv 0$ (within machine precision) would signal perfect convergence of the iteration, as observed in Ng-accelerated
numerical solutions of the O--Z equation in the forward direction \cite{Heinen2014}. For Ng-accelerated IHNCI, one observes
a rapid decay of $d_{i}(x)$ to small values at early stages of the iteration, but the function $d_{i}(x)$ never vanishes perfectly,
due to the statistical uncertainties in $g_{i}(x)$ and $S_{i}(y)$ that are being propagated to $\beta \mu_{i}(x)$.

The integral boundaries $x_{\text{min}}$ and $x_{\text{max}}$ in eq.~\eqref{eq:scalarprod} should be chosen such that most characteristic features of the
particle pair potential are included in the integration range, but $x_{\text{min}}$ should not be chosen too small:
The value of $x_{\text{min}}$ should be large enough to exclude the region where $\beta u_{i}(x) \gg 1$ from the integration range, because
otherwise all the calculated inner products will attain large values, rendering the iteration scheme rather insensitive to the small values
of $d_{i}(x)$ and $d_{l}^{(i)}(x)$ at larger values of $x$.
In all Ng-accelerated calculations that are presented here, we have chosen $x_{\text{min}} = 1$, and $x_{\text{max}}$ was
selected as the upper end of the interval on which $\beta u_{i}$ was sampled (typically, $x_{\text{max}} \approx 3$).

Note that the initial $(i = 1)$ step of the Ng iteration scheme is identical to a Picard iteration step, and that Ng-acceleration
can be applied to IBI as well IHNCI. The Ng-accelerated IBI algorithm is identical to the one described above, if only eq.~\eqref{eq:IHNCI_update}
is replaced by the equation $\beta \mu_{i}(x) = \beta u_{i}(x) + \ln\left[{g_i(x)}/{g_T(x)} \right]$. 
Deriving eqs.~\eqref{eq:IHNCI_update}--\eqref{eq:Ng_d_l^i_functions} is a straightforward task if Ng's original
instructions \cite{Ng1974} for the special case $i = 3$ are generalized to arbitrary values of the iteration index $i$
(see also ref. \cite{Heinen2014}). 

Iteration may be stopped when $\left( d_i(x), d_i(x) \right)$ is no longer significantly reducing as a function of $i$, but merely
fluctuating around a stationary mean. In the test cases that we have studied, convergence to this quasi-steady-state
occurs around $i = 10$ for both the Ng--accelerated IBI and IHNCI. Picard--IHNCI converges typically faster than Picard--IBI, which may
take hundreds of iterations until convergence is achieved \cite{Jain2006}.
\subsubsection*{\sffamily \large Monte Carlo simulation}\label{subsubsec:MC}
Every one of the (N,V,T) Metropolis MC simulations in the IHNCI algorithm uses a cubic simulation box of volume $V = L^3$ with
periodic boundary conditions in all three Cartesian directions and proceeds as follows:
In a first step, a dense disordered packing of monodisperse hard spheres is generated with the event-driven Lubachevsky-Stillinger
inflation algorithm \cite{Lubachevsky1990, Skoge2006}, running on a central processing unit (CPU).
The hard sphere packing fraction at which inflation is stopped is an adjustable parameter of the IHNCI method.
It should be chosen close to, or slightly higher than an estimated effective packing fraction of the particles that are being simulated.
After the initial inflation, the $N$-particle system is copied $M-1$ times, resulting in an ensemble of $M$ systems.
The following, statistically independent MC simulations of $M$ systems are carried out on a graphics processing unit (GPU) on $M$ cores.
In an equilibration phase, the particle interaction is first changed from the hard sphere no-overlap condition to the
reduced potential $\beta u_i(x)$. During equilibration, the maximum particle displacement is then dynamically adjusted
until either $\sim 50\%$ of all the (single particle, local) MC moves are accepted, or (in case of low-density systems)
until the maximal random displacement has reached the value $L/2$.
Equilibration is stopped after every particle has been moved $100$ times on average.
This simple ad-hoc criterion for the duration of the equilibration phase is justified in the cases reported here, by the observed
excellent agreement of the converged functions $g_i(x)$ and $S_i(y)$ with the input functions $g_T(x)$ and $S_T(y)$
(see fig.~\ref{fig:IBI_vs_IHNCI_LJ_T1_5_rho08} for a representative example).
Equilibration is followed by a MC production run, during which ensemble ergodicity is assumed and the correlation functions
$g_i(x)$ and $S_i(y)$ for each ensemble member are recorded on the GPU as averages over many particle configurations.
At the end of the production run, the correlation functions are ensemble-averaged on a CPU. 
\subsubsection*{\sffamily \large Correlation function extraction}\label{subsubsec:gx_Sy_extraction}
Extraction of $g_i(x)$ in the MC simulation is based on the straightforward eq.~\eqref{eq:gx} and requires no further comment.
Extraction of $S_i(y)$, on the other hand, requires special care: Since the function $S_i(y)$ is used in the IHNCI algorithm as
a source of information complementary to $g_i(x)$, it is not permissible to compute $S_i(y)$ via Fourier transformation of $g_i(x)$.
Instead, the structure factor must be computed directly from the particle coordinates, using the expression
\begin{equation}\label{eq:Sy}
S(y) = \dfrac{1}{N}
\left\langle
{\left[ \sum\limits_{j=1}^{N} \cos(\mathbf{q} \cdot \mathbf{r}_j) \right]}^2 +
{\left[ \sum\limits_{j=1}^{N} \sin(\mathbf{q} \cdot \mathbf{r}_j) \right]}^2
\right\rangle  
\end{equation}
or an equivalent expression that refers directly to the particle coordinates $\mathbf{r}_j$.
In eq.~\eqref{eq:Sy} the brackets $\left\langle \ldots \right\rangle$ indicate both an ergodic ensemble average, as in eq.~\eqref{eq:gx},
and a binned average over discrete wave vectors $\mathbf{q} = (2\pi / L) \left[ n_x \hat{\mathbf{e}}_x + n_y \hat{\mathbf{e}}_y + n_z \hat{\mathbf{e}}_z \right]$
that satisfy $y - \Delta_y /2 < \left|\mathbf{q}\right| n^{-1/3} \leq y + \Delta_y /2$ and that are commensurate with the periodic simulation box.
Here, $\Delta_y$ is a wavenumber bin width, $n_x$, $n_y$ and $n_z$ are positive integers, and $\hat{\mathbf{e}}_x$, $\hat{\mathbf{e}}_y$ and $\hat{\mathbf{e}}_z$
are the Cartesian unit vectors. Direct extraction of $S_i(y)$ via eq.~\eqref{eq:Sy} is equivalent to Fourier transformation of $g_i(y) - 1$ only in the thermodynamic
limit $V, N \to \infty$, carried out at constant $n$.

\begin{figure}
\centering
\includegraphics[width=.45\columnwidth]{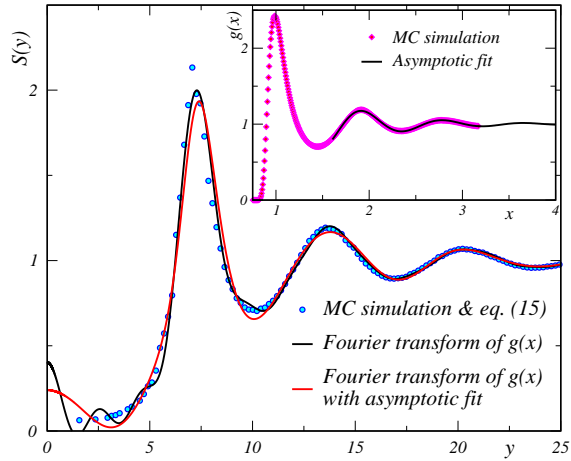}
\vspace{0em}
\caption{\label{fig:Sq_direct_and_from_gr_transform}
Structure factor (main panel) and radial distribution function (inset) for a Lennard-Jones fluid.
Symbols: Direct computation from particle coordinates, using eq.~\eqref{eq:gx} and eq.~\eqref{eq:Sy}.
Black and red curves in main panel: Structure factor approximations resulting from Fourier transformation
of the directly measured $g(x)$ without and with asymptotic fit extension, respectively.
Black curve in inset: Asymptotic fit extension of $g(x)$.
The Lennard-Jones parameters are the same as in fig.~\ref{fig:IBI_vs_IHNCI_LJ_T1_5_rho08}.
}
\end{figure}

For finite simulation boxes and finite numbers of particles, the expression in eq.~\eqref{eq:Sy}
provides additional information that is not contained in $g(x)$, as illustrated in fig.~\ref{fig:Sq_direct_and_from_gr_transform}:
The symbols in the figure inset represent a function $g(x)$ that was directly extracted for $x < 3.17$ from the coordinates of 256 Lennard-Jones particles in a MC simulation.
Approximating $g(x) = 0$ for $x > 3.17$ and Fourier-transforming the function $g(x) - 1$ results in the black curve in the main panel of
fig.~\ref{fig:Sq_direct_and_from_gr_transform}, which is a rather poor approximation of the structure factor that was directly calculated from the particle coordinates,
especially at small values of $y$. Function $S(y)$, directly extracted from the particle coordinates via eq.~\eqref{eq:Sy}, is represented by the 
symbols in the main panel of fig.~\ref{fig:Sq_direct_and_from_gr_transform}.
Assuming an asymptotic fit extension $g(x > 3.17) = 2.57 \times \exp\left\lbrace-1.40 x\right\rbrace \times \sin\left\lbrace7.23 x - 12.4\right\rbrace$
(black curve in the inset of fig.~\ref{fig:Sq_direct_and_from_gr_transform}) in the numerical Fourier transform in place of $g(x>3.17) = 0$ 
hardly results in any accuracy improvement, as can be seen from the resulting red curve in the main panel of fig.~\ref{fig:Sq_direct_and_from_gr_transform}.
The additional Fourier-space information from the directly calculated $S_i(y)$ and $S_T(y)$ is used in the IHNCI algorithm to generate a more precise iteration seed
and iteration update rule than those in the simpler IBI method, which is entirely based on the real-space information contained in $g_i(x)$ and $g_T(x)$.
Without displaying the results here, we have tested the IHNCI algorithm with input functions $g_T(x)$ and $S_T(y)$ where either $S_T(y)$ was
calculated as a numerical Fourier transform of $g_T(x)$ or vice versa. In either such case of improper use, IHNCI fails to converge to a reliable approximant of
the true particle pair potential.

The higher fidelity of IHNCI
(as compared to IBI, and demonstrated in the results section) comes at the price of higher computational complexity and less general applicability:
As a first disadvantage of IHNCI, the direct computation of $S_i(y)$ via eq.~\eqref{eq:Sy} represents the computational bottleneck of the method,
making a GPU implementation (or implementation on another massively parallel architecture) necessary to achieve acceptable runtimes of a few hours or less.
The second disadvantage of IHNCI is the requirement that the input (target) function $S_T(y)$ must not be obtained from Fourier transformation of the input $g_T(x)$,
for the same reasons as mentioned above, in relation to eq.~\eqref{eq:Sy}. Both functions $S_T(y)$ and $g_T(x)$ must result from direct evaluation of
expressions such as eqs.~\eqref{eq:Sy}, \eqref{eq:gx}, which require the particle coordinates $\mathbf{x}_i$ as inputs. The IHNCI method is therefore most
applicable for coarse graining in particle-based computer simulations, or for the analysis of experiments such as confocal microscopy \cite{Prasad2007} in which the particles
can be observed directly.

\section*{\sffamily \Large RESULTS}\label{sec:results}
The most solid indications for the supremacy of IHNCI over IBI are the magnitudes of the terms which are neglected in the respective central approximations:
While IHNCI makes an approximation of $\mathcal{O}(n^2)$ at the level of the bridge function, IBI imposes a stronger $\mathcal{O}(n)$ approximation at
the level of the PMF. 
Nonetheless, we are lacking a formal proof for a general advantage of IHNCI over IBI, and we must therefore resort to extensive, comparative testing
of both methods in the following.

We have generated input ('benchmark') functions $g_T(x)$, $S_T(y)$ for various test potentials $\beta u(x)$ by standard ('forward') MC simulation.
These simulations were essentially identical to the ones that are used within our implementation of IHNCI and IBI, with only one
principal difference: Instead of equilibration by $100$ moves per particle, the equilibration phase in the benchmark-generating MC simulations
was stopped only after reaching a more conservative thermalization criterion: It was required that the mean squared displacement of particles
be at least equal to $\sigma^2$ before equilibration was assumed and a production run was started.
Here, $\sigma$ is a (pseudo-)diameter of the particles, as defined separately for each test potential in the following.
The functions $g_T(x)$, $S_T(y)$ were then used as inputs for IBI and IHNCI. Any disagreement between the test potentials $\beta u(x)$
and the potentials $\lim_{i\to\infty} \beta u_i(x)$ to which IBI and IHNCI are converging quantifies an imperfection of the respective iterative method.  
In all cases where the L--J or SALR potentials were used as test potentials, we have used a cutoff radius $r_c = 5 \sigma$ for the potential in the benchmark-generating,
forward MC simulations. For particle separations $r > r_c$, the true potential was replaced by zero. The L--J and SALR acronyms are defined in the following.
\subsection*{\sffamily \large Test potentials}\label{subsec:TestPotentials}
As a first benchmark test, we use the two-parametric Lennard-Jones (L--J) 6--12 potential
\begin{equation}\label{eq:LJPot}
\beta u(r) =
4 \epsilon  \left[ {\left(\dfrac{\sigma}{r}\right)}^{12}  - {\left(\dfrac{\sigma}{r}\right)}^{6} \right],
\end{equation}
which depends on the prefactor $\epsilon$ and the soft particle diameter $\sigma$. 
Following the standard convention, the L--J potential can be characterized as well by
a reduced, dimensionless temperature $T^* = 1 / \epsilon$
and a reduced, dimensionless number density $\rho^* = n \sigma^3$.
Our second test case is the six-parametric Soft Steps potential
\begin{equation}\label{eq:SoftStepsPot}
\beta u(r) =
\dfrac{h_1}{\exp\left\lbrace {\left( \dfrac{r}{\sigma} - 1 \right)} / \delta_1 \right\rbrace + 1} +
\dfrac{h_2}{\exp\left\lbrace {\left( \dfrac{r}{\sigma} - (1 + w) \right)} / \delta_2 \right\rbrace + 1}
\end{equation}
in which $h_1 + h_2$ and $h_2$ are the heights of two rounded plateaus ('steps'), the first of which extends from
$r = 0$ to $r \approx \sigma$, and the second of which extends from $r \approx \sigma$ to $r \approx \sigma ( 1 + w )$.
In our simulations, we choose $h_1 \gg 1$ to prevent soft particle overlaps at $r \lesssim \sigma$.
The parameters $\delta_1$ and $\delta_2$ in eq.~\eqref{eq:SoftStepsPot} characterize the dimensionless skin depths of the
first and the second step.
We define $\phi = \pi \sigma^3 n / 6$ as a pseudo packing fraction
of the soft particles that interact via the potential in eq.~\eqref{eq:SoftStepsPot}.    
The third test case is the four-parametric Short-range Attraction, Long-range Repulsion (SALR) potential
\begin{equation}\label{eq:SALRPot}
\beta u(r) =
4 \epsilon  \left[ {\left(\dfrac{\sigma}{r}\right)}^{12}  - {\left(\dfrac{\sigma}{r}\right)}^{6} \right] +
A \dfrac{\sigma}{r} \exp\left\lbrace -\dfrac{r}{\sigma \xi} \right\rbrace
\end{equation}
which consists of a short-ranged L--J part and a long-ranged Yukawa part. For $\epsilon > 0$, the L--J part
exhibits short-ranged attraction as well as a strong repulsion at very short length scales $r \lesssim \sigma$,
which is not reflected in the SALR acronym. Instead, the Long-ranged Repulsion (LR) part of SALR   
refers to the Yukawa part of the potential with prefactor $A > 0$ and with a screening parameter $\xi > 0$.
As for the Soft Steps potential, we define $\phi = \pi \sigma^3 n / 6$ as a pseudo packing fraction
of the soft particles that interact via the potential in eq.~\eqref{eq:SALRPot}. 
Potentials of the SALR type are routinely used for modeling the interactions of proteins with competing
long-ranged electrostatic repulsion and short-ranged (van der Waals or hydration) attraction \cite{Godfrin2014, Riest2015, Das2018},
and are therefore representing a highly relevant application example.

A fourth test case is the three-parametric triangular potential
\begin{equation}\label{eq:TrianglePot}
\beta u(r) =
\left\lbrace
{
\begin{array}{ll}
0, & ~~\dfrac{r}{\sigma} > 1 + w \\~\\
\dfrac{2h\left(1 + w - \dfrac{r}{\sigma}\right)}{w}, & ~~1 + w \geq \dfrac{r}{\sigma} > 1 + \dfrac{w}{2} \\~\\
\dfrac{2h\left(\dfrac{r}{\sigma} - 1\right)}{w}, & ~~1 + \dfrac{w}{2} \geq \dfrac{r}{\sigma} > 1 \\~\\
\left|\dfrac{2h\left(\dfrac{r}{\sigma} - 1\right)}{w}\right|, & ~~1 \geq \dfrac{r}{\sigma}\\
\end{array}
}
\right.
\end{equation}
which is piecewise linear and continuous, consisting of a short-ranged repulsive part for $r < \sigma$
and a symmetric triangular part of height $h$ and reduced, dimensionless width $w$ between $r / \sigma = 1$ and $r / \sigma = 1 + w$.
In the simulations we cover both cases $h > 0$ and $h <  0$.  
As before, we define $\phi = \pi \sigma^3 n / 6$ as a pseudo packing fraction
of the soft particles that interact via the potential in eq.~\eqref{eq:TrianglePot}.    

Our list of test potentials concludes with the one-parametric pseudo hard sphere, shifted Mie 49--50 potential
\begin{equation}\label{eq:MiePot}
\beta u(r) =
\left\lbrace
{
\begin{array}{ll}
\dfrac{100}{3} {\left( \dfrac{50}{49} \right)}^{49} 
\left[ {\left( \dfrac{\sigma}{r} \right)}^{50} - {\left( \dfrac{\sigma}{r} \right)}^{49} \right] + \dfrac{2}{3}, & ~~ r < \left( \dfrac{50}{49} \right) \sigma \\~\\
0, & ~~r \geq \left( \dfrac{50}{49} \right) \sigma
\end{array}
}
\right.
\end{equation}
in which the prefactor has been tuned to achieve optimal agreement with the hard sphere equation of state \cite{Jover2012}.
Once again, we define $\phi = \pi \sigma^3 n / 6$ as a pseudo packing fraction
of the particles that interact via the potential in eq.~\eqref{eq:MiePot}.

\subsection*{\sffamily \large {Testing IHNCI vs. IBI: Picard iteration}}\label{subsec:TestingIBIvsIHNCI}
\begin{figure*}
\centering
\includegraphics[width=.8\textwidth]{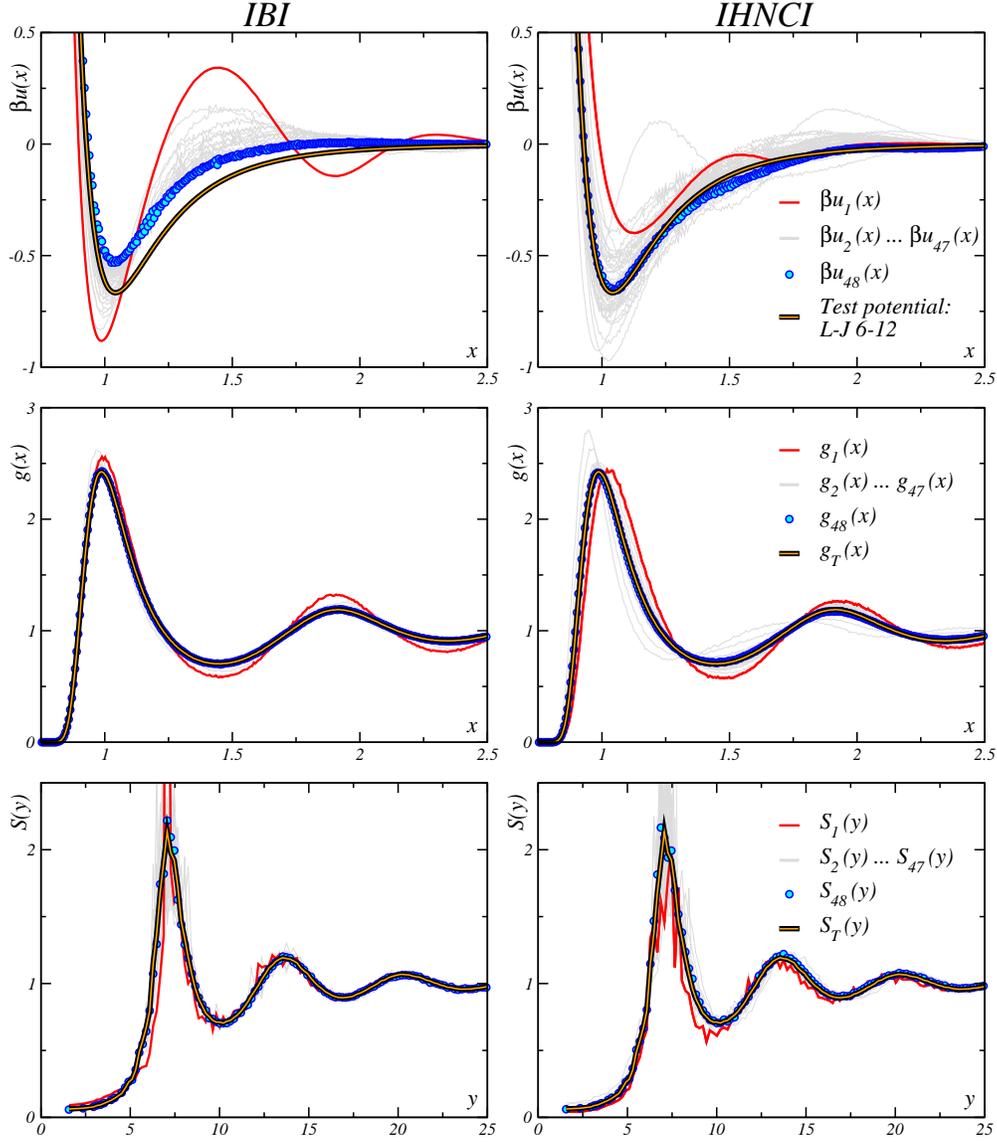}
\vspace{0em}
\caption{\label{fig:IBI_vs_IHNCI_LJ_T1_5_rho08}
Test of IBI (left column) and IHNCI (right column), both being used to compute a
reduced potential $\beta u(x)$ that recreates the target pair correlation functions $g_T(x)$ and $S_T(y)$
in the lower two rows of panels. The target correlation functions have been generated in a preceding MC simulation
with the L--J 6-12 test potential [eq.~\eqref{eq:LJPot}] that is represented by the black/orange curves in the top
row of panels. The L--J potential was cut off at $r_c = 5\sigma$.
Both IBI and IHNCI have been used in their simplest Picard-iteration forms and have been stopped after 48 iteration steps. 
The IHNCI method produces a function $\beta u_{48}(x)$ in significantly better agreement with the test
potential than the IBI method. The L--J parameters $T^* = 1.5$ and $\rho^* = 0.8$ correspond to
a rather dense, supercritical fluid \cite{Hansen1969, Lin2003}.      
}
\end{figure*}
Figure~\ref{fig:IBI_vs_IHNCI_LJ_T1_5_rho08} features the results of our IBI (left column of panels) and IHNCI (right column of panels)
calculations with identical input functions $g_T(x)$ and $S_T(y)$ that correspond to the L--J potential in eq.~\eqref{eq:LJPot}, for
$T^* = 1.5$ and $\rho^* = 0.8$.
Both IBI and IHNCI have been used here in their most simple Picard-iteration forms.
In this first test case, the IHNCI method is clearly surpassing the IBI method in terms of accuracy:

While the potential after 48 IBI steps (blue circles in the uppermost left column of fig.~\ref{fig:IBI_vs_IHNCI_LJ_T1_5_rho08})
deviates clearly from the test potential (black/orange curve) by more than $0.1~k_B T$ in an extended range of $x$ values,
the potential after 48 IHNCI steps (blue circles in the uppermost right column of fig.~\ref{fig:IBI_vs_IHNCI_LJ_T1_5_rho08})
is in good agreement with test potential, the principal difference being observed between $x \approx 1.25$ and $x \approx 1.75$,
where $\beta u_{48}(x)$ from IHNCI underestimates the test potential by up to $0.04~k_B T$.

Note that the iteration seeds
$\beta u_{1}(x)$ (red curves in the upper row of panels of fig.~\ref{fig:IBI_vs_IHNCI_LJ_T1_5_rho08}) for the two methods are distinctly different:
In case of IBI, $\beta u_{1}(x)$ equals the reduced PMF and deviates strongly from the target L--J potential and from the function $\beta u_{48}(x)$.
In the contrasting IHNCI case the function $\beta u_{1}(x)$, resulting from HNC inversion, is already rather close to the (practically) converged   
function $\beta u_{48}(x)$ which, in turn, is close to the L--J test potential.

In all the following test cases, we observe the same type of quality improvement in the iteration seed when switching
from IBI to IHNCI. Understanding IBI and IHNCI as functional minimization methods, the improvement of the iteration seed in IHNCI represents
a clear advantage that should generally help to avoid convergence into local minima. The second feature of IHNCI that helps to avoid trapping in
local, non-optimal minima is the improved iteration update rule in eq.~\eqref{eq:IHNCI_update}, as compared to eq.~\eqref{eq:IBI_update} for IBI.
In general, the IHNCI update rule should represent a better approximation of a true functional gradient descent that the simpler IBI update rule.

In the central and lower row of panels in fig.~\ref{fig:IBI_vs_IHNCI_LJ_T1_5_rho08}, we observe that both IBI and IHNCI are reproducing
the target functions $g_T(x)$ and $S_T(y)$ practically perfectly, within the level of the stochastic noise floor. This is a manifestation
of low practical usefulness of Henderson's theorem, as discussed in ref.~\cite{Potestio2014}: In spite of Henderson's theorem guaranteeing
the unique reversibility of the highly nonlinear functional mapping $\beta u(x) \to [g(x), S(y)]$ under ideal conditions, the practical implementation
of the reverse mapping $[g(x), S(y)] \to \beta u(x)$ is sincerely complicated in situations where $g(x), S(y)$ are only known within the range of a
statistical uncertainty. Many functions $\beta u(x)$ are reproducing a given pair of functions $g(x), S(y)$, even within small noise or uncertainty levels.
More than the IHNCI method, the IBI method is prone to error by converging onto just one of the many functions $\beta u(x)$ that will reproduce the
target pair correlation functions. Albeit not immune, the IHNCI method is significantly less prone to such failure as it uses a higher quality
iteration seed and a higher quality iteration update rule. The converged potential from IHNCI is thus more likely to be a good approximant of the true
particle pair potential.
For all test cases shown in the following, we have checked that both IBI and IHNCI are practically perfectly reproducing the input functions
$g_T(x)$ and $S_T(y)$. We therefore refrain from plotting the correlation functions in the remaining parts of this paper.  

\begin{figure}
\centering
\includegraphics[width=.3\columnwidth, angle=-90]{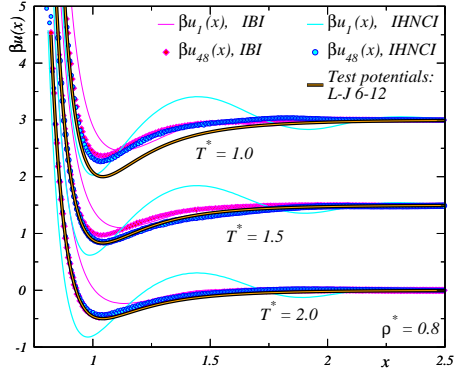}
\vspace{0em}
\caption{\label{fig:IBI_vs_IHNCI_LJ_3Temp}
Test of IBI (pink/red) and IHNCI (blue): Three L--J test potentials (black/orange curves) as defined in eq.~\eqref{eq:LJPot},
and with a cutoff radius of $r_c = 5\sigma$, have been used in MC simulations to compute the target correlation functions $g_T(x)$ and $S_T(y)$
which then served as an input to the IBI and IHNCI methods,
both of which were used in their simple Picard-iteration form.
The results in this figure are the reduced pair potentials
$\beta u_1(x)$ in the first iteration (solid blue and pink curves) and $\beta u_{48}(x)$, after 48 iterations (symbols)
in both the IBI and the IHNCI method.
All cases are for a reduced density $\rho^* = 0.8$. The three selected 
reduced temperatures $T^* = 2.0, 1.5$ and $1.0$ correspond to two supercritical fluid states and one liquid state, respectively \cite{Hansen1969, Lin2003}. 
}
\end{figure}
Let us now turn our attention to fig.~\ref{fig:IBI_vs_IHNCI_LJ_3Temp}, which features the potentials obtained from IBI (pink/red) and IHNCI (blue),
using three different L--J test potentials (black/orange curves) to be reproduced.
The reduced density for all test potentials in fig.~\ref{fig:IBI_vs_IHNCI_LJ_3Temp}
is $\rho^* = 0.8$, and the three different reduced temperatures $T^* = 2.0$, $T^* = 1.5$ (as in fig.~\ref{fig:IBI_vs_IHNCI_LJ_T1_5_rho08}) and $T^* = 1.0$ have been chosen,
corresponding to two supercritical fluid states and one liquid state \cite{Hansen1969, Lin2003}.
As a result, we observe a feature of IBI and IHNCI that appears universal among all test cases that we have studied, including some which are not featured in this paper
for the sake of brevity: At high temperatures, both IBI and IHNCI are quite
precise,
although some imperfection in the converged potentials may remain (see the lowermost
group of curves and symbols in fig.~\ref{fig:IBI_vs_IHNCI_LJ_3Temp}, for $T^* = 2.0$). When the temperature is gradually lowered, the IBI method fails first, while the
IHNCI method still retains a good
precision
(central group of curves and symbols in fig.~\ref{fig:IBI_vs_IHNCI_LJ_3Temp}, for $T^* = 1.5$). Eventually, at low temperatures
both IBI and IHNCI are failing to provide a reliable approximant of the test potential (uppermost group of curves and symbols in fig.~\ref{fig:IBI_vs_IHNCI_LJ_3Temp},
for $T^* = 1.0$).

A comment is in place here, regarding the thermodynamic inconsistency between the target fluids with correlation functions $g_T(x)$ and $S_T(y)$
and the fluids after 48 IBI or IHNCI iterations. Since a pressure (or ramp) correction \cite{Reith2003, Jain2006, Rosenberger2016}
has not been included in any of the IBI or IHNCI runs presented in this paper, the pressures of the target fluids are not matched in general by either of the two
iterative schemes. For the systems in fig.~\ref{fig:IBI_vs_IHNCI_LJ_3Temp}, we find reduced virial pressures of $\beta p / n = 3.59, 3.12$ and $1.81$ for the reference
L--J fluids at $T^* = 2.0, 1.5$ and $1.0$, respectively. The corresponding reduced virial pressures from the 48$^{\text{th}}$ MC simulation in the IBI algorithm are
$\beta p / n = 4.31, 4.38$ and $4.37$ while the respective 48$^{\text{th}}$ IHNCI iterations result in $\beta p / n = 4.90, 2.71$ and $4.14$.
Hence, both IBI and IHNCI show a similar degree of thermodynamic inconsistency with the target fluid, and both methods may profit from a ramp correction.
While a ramp correction could be easily included into IHNCI, we are not discussing it here for the sake of brevity and more general applicability to such
cases where the necessary thermodynamic information of the target fluid might not be available. Confocal microscopy of a suspension of particles is one example
where such thermodynamic information is hard to obtain within the accuracy that would be required.

\begin{figure}
\centering
\includegraphics[width=.3\columnwidth, angle=-90]{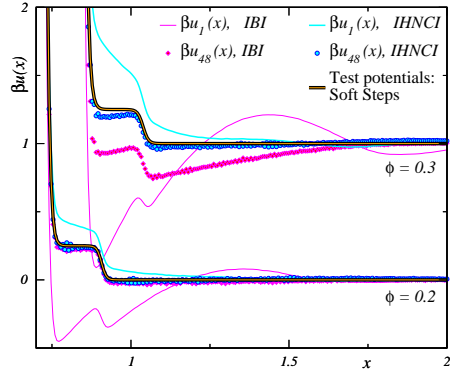}
\vspace{0em}
\caption{\label{fig:IBI_vs_IHNCI_SoftSteps_2Dens}
Same as fig.~\ref{fig:IBI_vs_IHNCI_LJ_3Temp}, but for two Soft Step test potentials as defined in eq.~\eqref{eq:SoftStepsPot}.
All cases are for $h_2 = 0.25, w = 0.25$ and $\delta_1 = \delta_2 = 10^{-2}$.
The lower group of results is for $\phi = 0.2, h_1 = 10$ and
the upper group of results is for $\phi = 0.3, h_1 = 50$.
}
\end{figure}
We turn now to fig.~\ref{fig:IBI_vs_IHNCI_SoftSteps_2Dens} in which the IBI and IHNCI results for the Soft Step test potential in eq.~\eqref{eq:SoftStepsPot}
are shown: One of the two test cases (the lower group of curves and symbols) is for a rather dilute system at $\phi = 0.2$, and the other case
is for a more crowded system at $\phi = 0.3$. While the converged potentials $\beta u_{48}(x)$ from both the IBI and the IHNCI method are in excellent agreement
with the test potential in case of $\phi = 0.2$, and the IHNCI method remains rather
precise
at $\phi = 0.3$, the IBI method fails drastically at that higher
density, converging to a potential that underestimates the test potential by more than $0.25~k_B T$.

\begin{figure}
\centering
\includegraphics[width=.3\columnwidth, angle=-90]{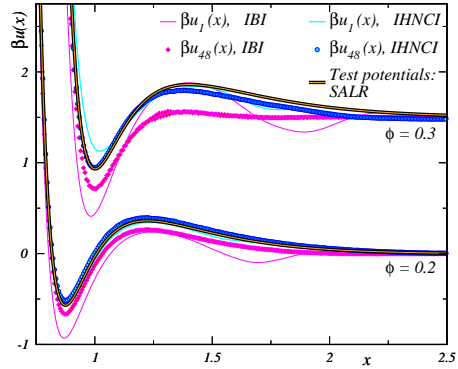}
\vspace{0em}
\caption{\label{fig:IBI_vs_IHNCI_SALR_2Dens}
Same as fig.~\ref{fig:IBI_vs_IHNCI_LJ_3Temp}, but for two SALR potentials as defined in eq.~\eqref{eq:SALRPot},
and cut off at $r_c = 5\sigma$.
All cases are for $\epsilon = 7, A = 75, \xi = 0.5$.
The lower group of results is for $\phi = 0.2$ and
the upper group of results is for $\phi = 0.3$.
}
\end{figure}
A qualitatively similar failure of IBI is revealed in fig.~\ref{fig:IBI_vs_IHNCI_SALR_2Dens},
which shows the results from two test cases with the SALR potential in eq.~\eqref{eq:SALRPot}:
At the lower SALR packing fraction $\phi = 0.2$, both IBI and IHNCI deliver quite accurate results,
but IHNCI is already significantly more
precise
than IBI even for this rather dilute system.
For the more concentrated SALR system at $\phi = 0.3$, IHNCI remains accurate while IBI fails drastically.

\begin{figure}
\centering
\includegraphics[width=.3\columnwidth, angle=-90]{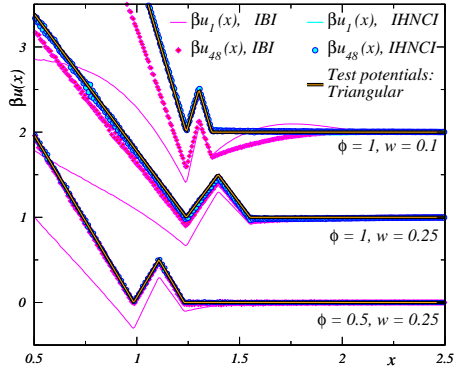}
\vspace{0em}
\caption{\label{fig:IBI_vs_IHNCI_RepRep_3Triangles}
Same as fig.~\ref{fig:IBI_vs_IHNCI_LJ_3Temp}, but for three triangular potentials as defined in eq.~\eqref{eq:TrianglePot}.
The blue solid curves for $\beta u_1(x)$ in the IHNCI method are not visible in the figure as they are overlapped by the
blue circles for $\beta u_{48}(x)$, IHNCI.
All cases are for $h = 0.5$.
The lower group of results is for $\phi = 0.5, w = 0.25$,
the central group is for $\phi = 1, w = 0.25$, and
the upper group of results is for $\phi = 1, w = 0.1$.
}
\end{figure}
\begin{figure}
\centering
\includegraphics[width=.3\columnwidth, angle=-90]{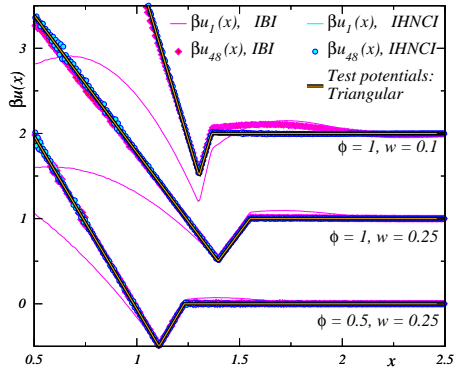}
\vspace{0em}
\caption{\label{fig:IBI_vs_IHNCI_RepAtt_3Triangles}
Same as fig.~\ref{fig:IBI_vs_IHNCI_RepRep_3Triangles}, but for $h = -0.5$.
}
\end{figure}

The observations made so far are confirmed by six additional test cases in which the triangular potential from eq.~\eqref{eq:TrianglePot} has been
used. The results for three purely repulsive triangular potentials (with $h > 0$) are shown in fig.~\ref{fig:IBI_vs_IHNCI_RepRep_3Triangles},
while fig.~\ref{fig:IBI_vs_IHNCI_RepAtt_3Triangles} features the results for three different triangular target potentials with short-ranged
repulsion that is followed by an attractive triangular well ($h < 0$). While both IBI and IHNCI are performing well for the test cases at lower density,
the IHNCI method proves to be superior for systems at higher density.

\begin{figure}
\centering
\includegraphics[width=.3\columnwidth, angle=-90]{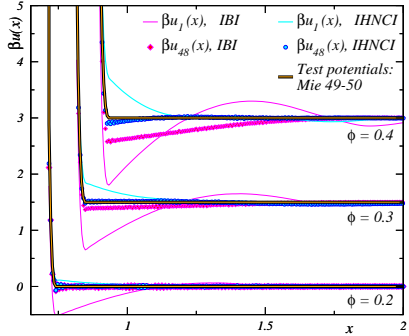}
\vspace{0em}
\caption{\label{fig:IBI_vs_IHNCI_Mie_3Dens}
Same as fig.~\ref{fig:IBI_vs_IHNCI_LJ_3Temp}, but for three Mie 49--50 pseudo hard sphere potentials as defined in eq.~\eqref{eq:MiePot}.
Results for three packing fractions $\phi = 0.2, 0.3$ and $0.4$ (from bottom to top) are shown. 
}
\end{figure}

Figure~\ref{fig:IBI_vs_IHNCI_Mie_3Dens} features the results from testing the capabilities of IBI and IHNCI to converge (close) to
the pseudo hard sphere, shifted Mie 49--50 potential in eq.~\eqref{eq:MiePot} at the three different packing fractions $\phi = 0.2, 0.3$ and $0.4$.
Once again the superiority of IHNCI over IBI is demonstrated, in particular for the most concentrated system at $\phi = 0.4$.

\subsection*{\sffamily \large Testing IHNCI vs. IBI: Ng iteration}\label{subsec:TestingNgIBIvsNgIHNCI}
All results presented so far have been obtained with the simple Picard-iteration versions of the IBI and IHNCI algorithms.
While IHNCI has proven to be more reliable in converging close to a known test potential, the numerical efficiency of the Picard iterations 
is less than satisfying due to the slow convergence of the algorithms, requiring hours of GPU time for a single run of 48 iterations.
Replacing Picard iteration by the more sophisticated Ng-accelerated iteration is therefore a promising approach towards the construction of
algorithms with higher practical applicability.

\begin{figure*}
\centering
\includegraphics[width=.8\textwidth]{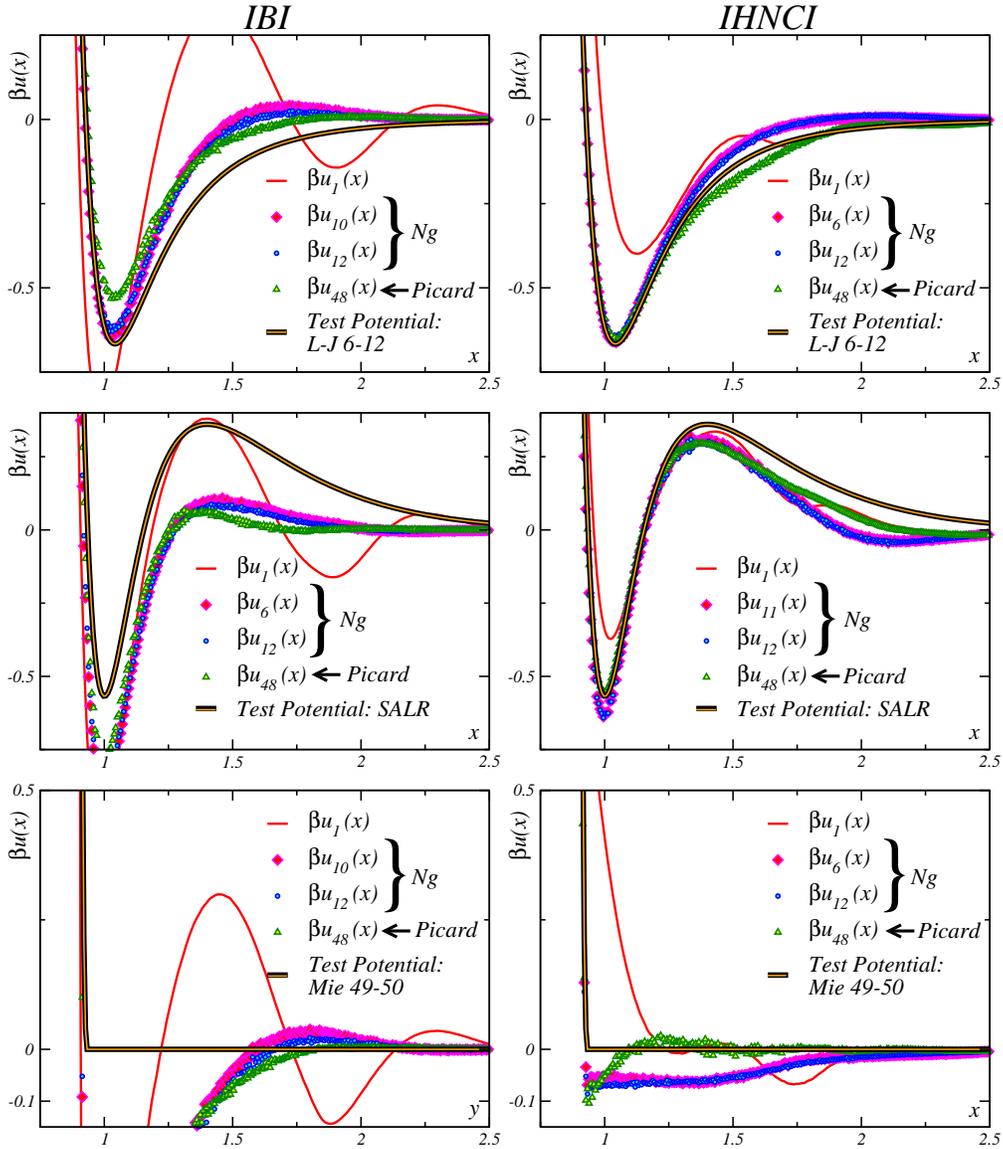}
\vspace{0em}
\caption{\label{fig:Ng-IBI_vs_Ng-IHNCI}
Test of the Ng-accelerated and standard Picard iteration versions of both the IBI (left column) and IHNCI (right column) method.
Three different test potentials of the L--J 6--12 (top row of panels), the SALR (center row of panels) and the Mie 49--50 (lower row of panels) type
are used as test cases. In all cases, the Ng accelerated iteration has converged after 12 iterations to a potential (blue circles) that is rather close,
but not identical to the potential which is obtained from the same seed $\beta u_1(x)$ (red solid curves) after 48 Picard iterations (green triangles).    
The L--J parameters are $T^* = 1.5$ and $\rho^* = 0.8$, while the SALR system is for $\epsilon = 7$, $A = 75$, $\xi = 0.5$, $\phi = 0.3$
and the Mie 49--50 system is for $\phi = 0.4$.      
}
\end{figure*}
In fig.~\ref{fig:Ng-IBI_vs_Ng-IHNCI} we test the performance of Ng-accelerated IBI and IHNCI versus the Picard iteration forms of the
two algorithms. Three different test potentials (L--J 6--12, SALR and shifted Mie 49--50) of markedly different shapes were chosen for these comparative tests.
The main result here is that both the Ng-accelerated IBI and IHNCI are converging significantly faster than their Picard iteration counterparts.
All Ng-accelerated calculations have converged after 12 iterations or less, including cases where convergence was observed after as little as 6
iterations. The potentials after 12 Picard iterations, which are significantly different from $\beta u_{48}(x)$ in all covered cases,
are not included in fig.~\ref{fig:Ng-IBI_vs_Ng-IHNCI} to avoid overcrowding the plot.  
It may be unexpected that the Ng-accelerated iterations are converging to potentials that are slightly, but significantly different from the
respective potentials obtained after 48 Picard iterations. Contrary to what one might presume, this is not because the pertinent Picard iterations
are far from convergence after 48 iterations: As we have tested, a small disagreement between the potentials from converged Ng-accelerated iterations and Picard iterations
remains even in cases where the temperature is high enough or the density is low enough to render the Picard iterations quickly convergent.
Hence, the functional fix point of Ng-iteration appears to be truly different from that of Picard iteration. This is consistent with the picture
in which the different iteration schemes represent functional gradient descent methods with different approximations of the true (unknown)
functional gradient. Each iteration method reaches a stationary (converged, modulo fluctuations) state when its approximation of the functional
gradient vanishes. For different approximations of the gradient, this happens at different states.  
For most practical purposes such as experimental data analysis or coarse graining in computer simulations,
the potentials $u_{12}(x)$ from the Ng-accelerated iterations should be as good (or bad) as the potentials that would be obtained 
after many more Picard iterations. 
Regarding computational complexity, let us remark that each of the IHNCI runs presented here was carried out with ensembles
of $M = 256$ systems, each containing $N = 256$ particles. To obtain rather good statistics for $S_{i}(y), i = 1, 2, \ldots 48$, a total
runtime of approximately $8$ hours was needed on an inexpensive GPU in case of Picard iteration.
The Ng-accelerated runs for $i = 1, 2, \ldots 12$ took around $4$ hours on the same machine.
A single iteration of the Ng-accelerated runs is slower than a single Picard iteration on average, because the Ng accelerated scheme
requires calculation of $g_{i}(x)$ and $S_{i}(y)$ with equal statistical uncertainty levels for all values of the iteration index $i$.
The reason for this is the sensitivity of eq.~\eqref{eq:Ng_iteration} to early iteration stages.
Picard iteration, on the other hand, is insensitive to its early stages when it approaches the stationary state.
The early stages of Picard iteration can therefore be carried out with less MC steps per simulation, and with a higher statistical
uncertainty in the measured functions $g_{i}(x)$ and $S_{i}(y)$ at small values of $i$. 
On a state-of-the art GPU it should be possible to reduce the typical time for one complete Ng-accelerated IHNCI run to $1$ hour or less.

\section*{\sffamily \Large CONCLUSIONS}
The novel IO--ZI method for the calculation of particle pair interaction potentials from given particle
pair correlations in disordered fluid phases has been presented in this work. Different O--Z closure relations
can be employed in conjunction with IO--ZI, but we have focused our attention here on the simplest case using
the HNC closure, resulting in the IHNCI method. 
The IHNCI (or, more generally, IO--ZI) method constitutes a generalization
of the established IBI method \cite{Reith2003} with generally higher
precision
and reliability.
In addition to the real space radial distribution function, the Fourier-space static structure factor is required
as an input for IO--ZI. This additional Fourier space information is the foundation of the improved
precision
of IO--ZI (as compared to IBI). At the same time, the requirement to work with Fourier space correlation functions
restricts the applicability of IO--ZI to such systems where the pertinent input data are available, and
it increases the computational complexity of IO--ZI significantly beyond that of IBI.
The Ng iteration scheme has been demonstrated to accelerate significantly the convergence of both IBI and IHNCI,
thereby improving the practical applicability of both algorithms. In cases where thermodynamic observables of the target
fluid are known, a ramp correction \cite{Reith2003, Jain2006, Rosenberger2016} can easily be added to both IBI and IHNCI,
and this should generally result in an improvement of both methods.

Many of our test cases reveal a good performance of the simple IBI method.
If the fluid density is sufficiently low, or the temperature sufficiently high, IBI converges to pair
potentials that are in very good agreement with the test potentials.
This is related to the fact that the PMF $w(r)$ reduces to the pair potential $u(r)$ in the limit of small densities. 
Nevertheless, IBI fails to converge close to the test potentials when the density is increased
or the temperature is lowered, and these failures of IBI set in significantly earlier than the
corresponding IHNCI failures among all of our test cases.
As a consequence, we propose to use IHNCI whenever the available input data allows to do so, and
whenever there is a doubt about the accuracy of IBI.

The IHNCI code is available from the author upon request.

\subsection*{\sffamily \large ACKNOWLEDGMENTS}
It is my pleasure to thank Ram\'{o}n Casta\~n{e}da Priego and Marco Laurati for our many discussions
related to the present work, and Raffaello Potestio for his criticism on the practical usefulness of Henderson's theorem,
which has triggered my interest in this topic.
I acknowledge financial support from CONACyT (Grant No. 237425/2014).


\clearpage



\begin{thebibliography}{45}
\expandafter\ifx\csname natexlab\endcsname\relax\def\natexlab#1{#1}\fi
\expandafter\ifx\csname bibnamefont\endcsname\relax
  \def\bibnamefont#1{#1}\fi
\expandafter\ifx\csname bibfnamefont\endcsname\relax
  \def\bibfnamefont#1{#1}\fi
\expandafter\ifx\csname citenamefont\endcsname\relax
  \def\citenamefont#1{#1}\fi
\expandafter\ifx\csname url\endcsname\relax
  \def\url#1{\texttt{#1}}\fi
\expandafter\ifx\csname urlprefix\endcsname\relax\def\urlprefix{URL }\fi
\providecommand{\bibinfo}[2]{#2}
\providecommand{\eprint}[2][]{\url{#2}}

\bibitem[{\citenamefont{Henderson}(1974)}]{Henderson1974}
\bibinfo{author}{\bibfnamefont{R.~L.} \bibnamefont{Henderson}},
  \bibinfo{journal}{Phys. Lett.} \textbf{\bibinfo{volume}{A49}},
  \bibinfo{pages}{197} (\bibinfo{year}{1974}).

\bibitem[{\citenamefont{Potestio et~al.}(2014)\citenamefont{Potestio, Peter,
  and Kremer}}]{Potestio2014}
\bibinfo{author}{\bibfnamefont{R.}~\bibnamefont{Potestio}},
  \bibinfo{author}{\bibfnamefont{C.}~\bibnamefont{Peter}}, \bibnamefont{and}
  \bibinfo{author}{\bibfnamefont{K.}~\bibnamefont{Kremer}},
  \bibinfo{journal}{Entropy} \textbf{\bibinfo{volume}{16}},
  \bibinfo{pages}{4199} (\bibinfo{year}{2014}).

\bibitem[{\citenamefont{Reith et~al.}(2003)\citenamefont{Reith, P\"{u}tz, and
  M\"{u}ller-Plathe}}]{Reith2003}
\bibinfo{author}{\bibfnamefont{D.}~\bibnamefont{Reith}},
  \bibinfo{author}{\bibfnamefont{M.}~\bibnamefont{P\"{u}tz}}, \bibnamefont{and}
  \bibinfo{author}{\bibfnamefont{F.}~\bibnamefont{M\"{u}ller-Plathe}},
  \bibinfo{journal}{J. Comput. Chem.} \textbf{\bibinfo{volume}{24}},
  \bibinfo{pages}{1624} (\bibinfo{year}{2003}).

\bibitem[{\citenamefont{Jain et~al.}(2006)\citenamefont{Jain, Garde, and
  Kumar}}]{Jain2006}
\bibinfo{author}{\bibfnamefont{S.}~\bibnamefont{Jain}},
  \bibinfo{author}{\bibfnamefont{S.}~\bibnamefont{Garde}}, \bibnamefont{and}
  \bibinfo{author}{\bibfnamefont{S.~K.} \bibnamefont{Kumar}},
  \bibinfo{journal}{Ind. Eng. Chem. Res.} \textbf{\bibinfo{volume}{45}},
  \bibinfo{pages}{5614} (\bibinfo{year}{2006}).

\bibitem[{\citenamefont{Moore et~al.}(2014)\citenamefont{Moore, Iacovella, and
  McCabe}}]{Moore2014}
\bibinfo{author}{\bibfnamefont{T.~C.} \bibnamefont{Moore}},
  \bibinfo{author}{\bibfnamefont{C.~R.} \bibnamefont{Iacovella}},
  \bibnamefont{and} \bibinfo{author}{\bibfnamefont{C.}~\bibnamefont{McCabe}},
  \bibinfo{journal}{J. Chem. Phys.} \textbf{\bibinfo{volume}{140}},
  \bibinfo{pages}{224104} (\bibinfo{year}{2014}).

\bibitem[{\citenamefont{Dijkstra et~al.}(2000)\citenamefont{Dijkstra, van Roij,
  and Evans}}]{Dijkstra2000}
\bibinfo{author}{\bibfnamefont{M.}~\bibnamefont{Dijkstra}},
  \bibinfo{author}{\bibfnamefont{R.}~\bibnamefont{van Roij}}, \bibnamefont{and}
  \bibinfo{author}{\bibfnamefont{R.}~\bibnamefont{Evans}}, \bibinfo{journal}{J.
  Chem. Phys.} \textbf{\bibinfo{volume}{113}}, \bibinfo{pages}{4799}
  (\bibinfo{year}{2000}).

\bibitem[{\citenamefont{Louis}(2001)}]{Louis2001}
\bibinfo{author}{\bibfnamefont{A.~A.} \bibnamefont{Louis}},
  \bibinfo{journal}{Phil. Trans. R. Soc. Lond. A}
  \textbf{\bibinfo{volume}{359}}, \bibinfo{pages}{939} (\bibinfo{year}{2001}).

\bibitem[{\citenamefont{Dobnikar et~al.}(2006)\citenamefont{Dobnikar,
  Casta\~{n}eda Priego, von Gr\"{u}nberg, and Trizac}}]{Dobnikar2006}
\bibinfo{author}{\bibfnamefont{J.}~\bibnamefont{Dobnikar}},
  \bibinfo{author}{\bibfnamefont{R.}~\bibnamefont{Casta\~{n}eda Priego}},
  \bibinfo{author}{\bibfnamefont{H.~H.} \bibnamefont{von Gr\"{u}nberg}},
  \bibnamefont{and} \bibinfo{author}{\bibfnamefont{E.}~\bibnamefont{Trizac}},
  \bibinfo{journal}{New J. of Phys.} \textbf{\bibinfo{volume}{8}},
  \bibinfo{pages}{277} (\bibinfo{year}{2006}).

\bibitem[{\citenamefont{Trizac et~al.}(2007)\citenamefont{Trizac, Belloni,
  Dobnikar, von Gr\"{u}nberg, and Casta\~{n}eda Priego}}]{Trizac2007}
\bibinfo{author}{\bibfnamefont{E.}~\bibnamefont{Trizac}},
  \bibinfo{author}{\bibfnamefont{L.}~\bibnamefont{Belloni}},
  \bibinfo{author}{\bibfnamefont{J.}~\bibnamefont{Dobnikar}},
  \bibinfo{author}{\bibfnamefont{H.~H.} \bibnamefont{von Gr\"{u}nberg}},
  \bibnamefont{and}
  \bibinfo{author}{\bibfnamefont{R.}~\bibnamefont{Casta\~{n}eda Priego}},
  \bibinfo{journal}{Phys. Rev. E} \textbf{\bibinfo{volume}{76}},
  \bibinfo{pages}{011401} (\bibinfo{year}{2007}).

\bibitem[{\citenamefont{Heinen et~al.}(2014{\natexlab{a}})\citenamefont{Heinen,
  Palberg, and L\"{o}wen}}]{HeinenPalbergLoewen2014}
\bibinfo{author}{\bibfnamefont{M.}~\bibnamefont{Heinen}},
  \bibinfo{author}{\bibfnamefont{T.}~\bibnamefont{Palberg}}, \bibnamefont{and}
  \bibinfo{author}{\bibfnamefont{H.}~\bibnamefont{L\"{o}wen}},
  \bibinfo{journal}{J. Chem. Phys.} \textbf{\bibinfo{volume}{140}},
  \bibinfo{pages}{124904} (\bibinfo{year}{2014}{\natexlab{a}}).

\bibitem[{\citenamefont{Rosenberger et~al.}(2016)\citenamefont{Rosenberger,
  Hanke, and van~der Vegt}}]{Rosenberger2016}
\bibinfo{author}{\bibfnamefont{D.}~\bibnamefont{Rosenberger}},
  \bibinfo{author}{\bibfnamefont{M.}~\bibnamefont{Hanke}}, \bibnamefont{and}
  \bibinfo{author}{\bibfnamefont{N.~F.~A.} \bibnamefont{van~der Vegt}},
  \bibinfo{journal}{Eur. Phys. J.-Spec. Top.} \textbf{\bibinfo{volume}{225}},
  \bibinfo{pages}{1323} (\bibinfo{year}{2016}).

\bibitem[{\citenamefont{Hansen and McDonald}(1986)}]{Hansen_McDonald1986}
\bibinfo{author}{\bibfnamefont{J.-P.} \bibnamefont{Hansen}} \bibnamefont{and}
  \bibinfo{author}{\bibfnamefont{I.~R.} \bibnamefont{McDonald}},
  \emph{\bibinfo{title}{Theory of Simple Liquids}}
  (\bibinfo{publisher}{Academic Press}, \bibinfo{address}{London},
  \bibinfo{year}{1986}), \bibinfo{edition}{2nd} ed.

\bibitem[{\citenamefont{Morita}(1958)}]{Morita1958}
\bibinfo{author}{\bibfnamefont{T.}~\bibnamefont{Morita}},
  \bibinfo{journal}{Prog. Theo. Phys.} \textbf{\bibinfo{volume}{20}},
  \bibinfo{pages}{920} (\bibinfo{year}{1958}).

\bibitem[{\citenamefont{Ng}(1974)}]{Ng1974}
\bibinfo{author}{\bibfnamefont{K.-C.} \bibnamefont{Ng}}, \bibinfo{journal}{J.
  Chem. Phys.} \textbf{\bibinfo{volume}{61}}, \bibinfo{pages}{2680}
  (\bibinfo{year}{1974}).

\bibitem[{\citenamefont{Pulay}(1980)}]{Pulay1980}
\bibinfo{author}{\bibfnamefont{P.}~\bibnamefont{Pulay}},
  \bibinfo{journal}{Chem. Phys. Lett.} \textbf{\bibinfo{volume}{73}},
  \bibinfo{pages}{393} (\bibinfo{year}{1980}).

\bibitem[{\citenamefont{Pulay}(1982)}]{Pulay1982}
\bibinfo{author}{\bibfnamefont{P.}~\bibnamefont{Pulay}}, \bibinfo{journal}{J.
  Comput. Chem.} \textbf{\bibinfo{volume}{3}}, \bibinfo{pages}{556}
  (\bibinfo{year}{1982}).

\bibitem[{\citenamefont{Heinen et~al.}(2014{\natexlab{b}})\citenamefont{Heinen,
  Allahyarov, and L\"{o}wen}}]{Heinen2014}
\bibinfo{author}{\bibfnamefont{M.}~\bibnamefont{Heinen}},
  \bibinfo{author}{\bibfnamefont{E.}~\bibnamefont{Allahyarov}},
  \bibnamefont{and}
  \bibinfo{author}{\bibfnamefont{H.}~\bibnamefont{L\"{o}wen}},
  \bibinfo{journal}{J. Comput. Chem.} \textbf{\bibinfo{volume}{35}},
  \bibinfo{pages}{275} (\bibinfo{year}{2014}{\natexlab{b}}).

\bibitem[{\citenamefont{Fries and Patey}(1985)}]{Fries1985}
\bibinfo{author}{\bibfnamefont{P.~H.} \bibnamefont{Fries}} \bibnamefont{and}
  \bibinfo{author}{\bibfnamefont{G.~N.} \bibnamefont{Patey}},
  \bibinfo{journal}{J. Chem. Phys.} \textbf{\bibinfo{volume}{82}},
  \bibinfo{pages}{429} (\bibinfo{year}{1985}).

\bibitem[{\citenamefont{Plischke and Henderson}(1986)}]{Plischke1986}
\bibinfo{author}{\bibfnamefont{M.}~\bibnamefont{Plischke}} \bibnamefont{and}
  \bibinfo{author}{\bibfnamefont{D.}~\bibnamefont{Henderson}},
  \bibinfo{journal}{J. Chem. Phys.} \textbf{\bibinfo{volume}{84}},
  \bibinfo{pages}{2846} (\bibinfo{year}{1986}).

\bibitem[{\citenamefont{Kjellander and Sarman}(1991)}]{Kjellander1991}
\bibinfo{author}{\bibfnamefont{R.}~\bibnamefont{Kjellander}} \bibnamefont{and}
  \bibinfo{author}{\bibfnamefont{S.}~\bibnamefont{Sarman}},
  \bibinfo{journal}{Mol. Phys.} \textbf{\bibinfo{volume}{74}},
  \bibinfo{pages}{665} (\bibinfo{year}{1991}).

\bibitem[{\citenamefont{Nyg\aa{}rd et~al.}(2012)\citenamefont{Nyg\aa{}rd,
  Kjellander, Sarman, Chodankar, Perret, Buitenhuis, and van~der
  Veen}}]{Nygard2012}
\bibinfo{author}{\bibfnamefont{K.}~\bibnamefont{Nyg\aa{}rd}},
  \bibinfo{author}{\bibfnamefont{R.}~\bibnamefont{Kjellander}},
  \bibinfo{author}{\bibfnamefont{S.}~\bibnamefont{Sarman}},
  \bibinfo{author}{\bibfnamefont{S.}~\bibnamefont{Chodankar}},
  \bibinfo{author}{\bibfnamefont{E.}~\bibnamefont{Perret}},
  \bibinfo{author}{\bibfnamefont{J.}~\bibnamefont{Buitenhuis}},
  \bibnamefont{and} \bibinfo{author}{\bibfnamefont{J.~F.} \bibnamefont{van~der
  Veen}}, \bibinfo{journal}{Phys. Rev. Lett.} \textbf{\bibinfo{volume}{108}},
  \bibinfo{pages}{037802} (\bibinfo{year}{2012}).

\bibitem[{\citenamefont{Ch\'{a}vez-P\'{a}ez
  et~al.}(2003)\citenamefont{Ch\'{a}vez-P\'{a}ez, Gonz\'{a}lez-Mozuelos,
  Medina-Noyola, and M\'{e}ndez-Alcaraz}}]{Chavez-Paez2003}
\bibinfo{author}{\bibfnamefont{M.}~\bibnamefont{Ch\'{a}vez-P\'{a}ez}},
  \bibinfo{author}{\bibfnamefont{P.}~\bibnamefont{Gonz\'{a}lez-Mozuelos}},
  \bibinfo{author}{\bibfnamefont{M.}~\bibnamefont{Medina-Noyola}},
  \bibnamefont{and} \bibinfo{author}{\bibfnamefont{J.~M.}
  \bibnamefont{M\'{e}ndez-Alcaraz}}, \bibinfo{journal}{J. Chem. Phys.}
  \textbf{\bibinfo{volume}{119}}, \bibinfo{pages}{7461} (\bibinfo{year}{2003}).

\bibitem[{\citenamefont{Contreras-Aburto
  et~al.}(2010)\citenamefont{Contreras-Aburto, M\'{e}ndez-Alcaraz, and
  Casta\~{n}eda Priego}}]{Contreras-Aburto2010}
\bibinfo{author}{\bibfnamefont{C.}~\bibnamefont{Contreras-Aburto}},
  \bibinfo{author}{\bibfnamefont{J.~M.} \bibnamefont{M\'{e}ndez-Alcaraz}},
  \bibnamefont{and}
  \bibinfo{author}{\bibfnamefont{R.}~\bibnamefont{Casta\~{n}eda Priego}},
  \bibinfo{journal}{J. Chem. Phys.} \textbf{\bibinfo{volume}{132}},
  \bibinfo{pages}{174111} (\bibinfo{year}{2010}).

\bibitem[{\citenamefont{Jaiswal et~al.}(2014)\citenamefont{Jaiswal, Bharadwaj,
  and Singh}}]{Jaiswal2014}
\bibinfo{author}{\bibfnamefont{A.}~\bibnamefont{Jaiswal}},
  \bibinfo{author}{\bibfnamefont{A.~S.} \bibnamefont{Bharadwaj}},
  \bibnamefont{and} \bibinfo{author}{\bibfnamefont{Y.}~\bibnamefont{Singh}},
  \bibinfo{journal}{J. Chem. Phys.} \textbf{\bibinfo{volume}{140}},
  \bibinfo{pages}{211103} (\bibinfo{year}{2014}).

\bibitem[{\citenamefont{Mendoza-M\'{e}ndez
  et~al.}(2017)\citenamefont{Mendoza-M\'{e}ndez, L\'{a}zaro-L\'{a}zaro,
  S\'{a}nchez-D\'{i}az, Ram\'{i}rez-Gonz\'{a}lez, P\'{e}rez-\'{A}ngel, and
  Medina-Noyola}}]{MendozaMendez2017}
\bibinfo{author}{\bibfnamefont{P.}~\bibnamefont{Mendoza-M\'{e}ndez}},
  \bibinfo{author}{\bibfnamefont{E.}~\bibnamefont{L\'{a}zaro-L\'{a}zaro}},
  \bibinfo{author}{\bibfnamefont{L.~E.} \bibnamefont{S\'{a}nchez-D\'{i}az}},
  \bibinfo{author}{\bibfnamefont{P.~E.}
  \bibnamefont{Ram\'{i}rez-Gonz\'{a}lez}},
  \bibinfo{author}{\bibfnamefont{G.}~\bibnamefont{P\'{e}rez-\'{A}ngel}},
  \bibnamefont{and}
  \bibinfo{author}{\bibfnamefont{M.}~\bibnamefont{Medina-Noyola}},
  \bibinfo{journal}{Phys. Rev. E} \textbf{\bibinfo{volume}{96}},
  \bibinfo{pages}{022608} (\bibinfo{year}{2017}).

\bibitem[{\citenamefont{G\"{o}tze}(2009)}]{Goetze2009}
\bibinfo{author}{\bibfnamefont{W.}~\bibnamefont{G\"{o}tze}},
  \emph{\bibinfo{title}{Complex Dynamics of Glass-Forming Liquids: A
  Mode-Coupling Theory}} (\bibinfo{publisher}{Oxford University Press},
  \bibinfo{address}{Oxford}, \bibinfo{year}{2009}).

\bibitem[{\citenamefont{N\"{a}gele}(1996)}]{Nagele1996}
\bibinfo{author}{\bibfnamefont{G.}~\bibnamefont{N\"{a}gele}},
  \bibinfo{journal}{Phys. Rep.} \textbf{\bibinfo{volume}{272}},
  \bibinfo{pages}{216} (\bibinfo{year}{1996}).

\bibitem[{\citenamefont{Heinen et~al.}(2011)\citenamefont{Heinen, Holmqvist,
  Banchio, and N\"{a}gele}}]{Heinen2011}
\bibinfo{author}{\bibfnamefont{M.}~\bibnamefont{Heinen}},
  \bibinfo{author}{\bibfnamefont{P.}~\bibnamefont{Holmqvist}},
  \bibinfo{author}{\bibfnamefont{A.~J.} \bibnamefont{Banchio}},
  \bibnamefont{and}
  \bibinfo{author}{\bibfnamefont{G.}~\bibnamefont{N\"{a}gele}},
  \bibinfo{journal}{J. Chem. Phys.} \textbf{\bibinfo{volume}{134}},
  \bibinfo{pages}{044532, \textit{ibid.} 129901} (\bibinfo{year}{2011}).

\bibitem[{Hei()}]{Heinen2011err}
\bibinfo{note}{Note that a factor $1/n$ is missing in equation [A1] of
  ref~\cite{Heinen2011}}.

\bibitem[{\citenamefont{Prasad et~al.}(2007)\citenamefont{Prasad, Semwogerere,
  and Weeks}}]{Prasad2007}
\bibinfo{author}{\bibfnamefont{V.}~\bibnamefont{Prasad}},
  \bibinfo{author}{\bibfnamefont{D.}~\bibnamefont{Semwogerere}},
  \bibnamefont{and} \bibinfo{author}{\bibfnamefont{E.~R.} \bibnamefont{Weeks}},
  \bibinfo{journal}{J. Phys.: Condens. Matter} \textbf{\bibinfo{volume}{19}},
  \bibinfo{pages}{113102} (\bibinfo{year}{2007}).

\bibitem[{\citenamefont{Percus and Yevick}(1958)}]{Percus1958}
\bibinfo{author}{\bibfnamefont{J.~K.} \bibnamefont{Percus}} \bibnamefont{and}
  \bibinfo{author}{\bibfnamefont{G.~J.} \bibnamefont{Yevick}},
  \bibinfo{journal}{Phys. Rev.} \textbf{\bibinfo{volume}{110}},
  \bibinfo{pages}{1} (\bibinfo{year}{1958}).

\bibitem[{\citenamefont{Kinoshita}(2003)}]{Kinoshita2003}
\bibinfo{author}{\bibfnamefont{M.}~\bibnamefont{Kinoshita}},
  \bibinfo{journal}{J. Chem. Phys.} \textbf{\bibinfo{volume}{118}},
  \bibinfo{pages}{8969} (\bibinfo{year}{2003}).

\bibitem[{\citenamefont{Hamilton}(2000)}]{Hamilton2000}
\bibinfo{author}{\bibfnamefont{A.~J.~S.} \bibnamefont{Hamilton}},
  \bibinfo{journal}{Mon. Not. R. Astron. Soc.} \textbf{\bibinfo{volume}{312}},
  \bibinfo{pages}{257} (\bibinfo{year}{2000}).

\bibitem[{Ham()}]{Hamilton_website}
\emph{\bibinfo{title}{\mbox{A.~J.~S.~Hamilton's FFTLog website:}}},
  \bibinfo{note}{\url{http://casa.colorado.edu/$\sim$ajsh/FFTLog/}}.

\bibitem[{\citenamefont{Talman}(1978)}]{Talman1978}
\bibinfo{author}{\bibfnamefont{J.~D.} \bibnamefont{Talman}},
  \bibinfo{journal}{J. Comput. Phys.} \textbf{\bibinfo{volume}{29}},
  \bibinfo{pages}{35} (\bibinfo{year}{1978}).

\bibitem[{\citenamefont{Heinen et~al.}(2015{\natexlab{a}})\citenamefont{Heinen,
  Horbach, and L\"{o}wen}}]{Heinen2015}
\bibinfo{author}{\bibfnamefont{M.}~\bibnamefont{Heinen}},
  \bibinfo{author}{\bibfnamefont{J.}~\bibnamefont{Horbach}}, \bibnamefont{and}
  \bibinfo{author}{\bibfnamefont{H.}~\bibnamefont{L\"{o}wen}},
  \bibinfo{journal}{Mol. Phys.} \textbf{\bibinfo{volume}{113}},
  \bibinfo{pages}{1164} (\bibinfo{year}{2015}{\natexlab{a}}).

\bibitem[{\citenamefont{Heinen et~al.}(2015{\natexlab{b}})\citenamefont{Heinen,
  Schnyder, Brady, and L\"{o}wen}}]{Heinen2015PRL}
\bibinfo{author}{\bibfnamefont{M.}~\bibnamefont{Heinen}},
  \bibinfo{author}{\bibfnamefont{S.~K.} \bibnamefont{Schnyder}},
  \bibinfo{author}{\bibfnamefont{J.~F.} \bibnamefont{Brady}}, \bibnamefont{and}
  \bibinfo{author}{\bibfnamefont{H.}~\bibnamefont{L\"{o}wen}},
  \bibinfo{journal}{Phys. Rev. Lett.} \textbf{\bibinfo{volume}{115}},
  \bibinfo{pages}{097801} (\bibinfo{year}{2015}{\natexlab{b}}).

\bibitem[{\citenamefont{Lubachevsky and Stillinger}(1990)}]{Lubachevsky1990}
\bibinfo{author}{\bibfnamefont{B.~D.} \bibnamefont{Lubachevsky}}
  \bibnamefont{and} \bibinfo{author}{\bibfnamefont{F.~H.}
  \bibnamefont{Stillinger}}, \bibinfo{journal}{J. Stat. Phys.}
  \textbf{\bibinfo{volume}{60}}, \bibinfo{pages}{561 } (\bibinfo{year}{1990}).

\bibitem[{\citenamefont{Skoge et~al.}(2006)\citenamefont{Skoge, Donev,
  Stillinger, and Torquato}}]{Skoge2006}
\bibinfo{author}{\bibfnamefont{M.}~\bibnamefont{Skoge}},
  \bibinfo{author}{\bibfnamefont{A.}~\bibnamefont{Donev}},
  \bibinfo{author}{\bibfnamefont{F.~H.} \bibnamefont{Stillinger}},
  \bibnamefont{and} \bibinfo{author}{\bibfnamefont{S.}~\bibnamefont{Torquato}},
  \bibinfo{journal}{Phys. Rev. E} \textbf{\bibinfo{volume}{74}},
  \bibinfo{pages}{041127} (\bibinfo{year}{2006}).

\bibitem[{\citenamefont{Godfrin et~al.}(2014)\citenamefont{Godfrin,
  Valadez-P\'{e}rez, Casta\~{n}eda Priego, Wagner, and Liu}}]{Godfrin2014}
\bibinfo{author}{\bibfnamefont{P.~D.} \bibnamefont{Godfrin}},
  \bibinfo{author}{\bibfnamefont{N.~E.} \bibnamefont{Valadez-P\'{e}rez}},
  \bibinfo{author}{\bibfnamefont{R.}~\bibnamefont{Casta\~{n}eda Priego}},
  \bibinfo{author}{\bibfnamefont{N.~J.} \bibnamefont{Wagner}},
  \bibnamefont{and} \bibinfo{author}{\bibfnamefont{Y.}~\bibnamefont{Liu}},
  \bibinfo{journal}{Soft Matter} \textbf{\bibinfo{volume}{10}},
  \bibinfo{pages}{5061} (\bibinfo{year}{2014}).

\bibitem[{\citenamefont{Riest and N\"{a}gele}(2015)}]{Riest2015}
\bibinfo{author}{\bibfnamefont{J.}~\bibnamefont{Riest}} \bibnamefont{and}
  \bibinfo{author}{\bibfnamefont{G.}~\bibnamefont{N\"{a}gele}},
  \bibinfo{journal}{Soft Matter} \textbf{\bibinfo{volume}{11}},
  \bibinfo{pages}{9273} (\bibinfo{year}{2015}).

\bibitem[{\citenamefont{Das et~al.}(2018)\citenamefont{Das, Riest, Winkler,
  Gompper, Dhont, and N\"{a}gele}}]{Das2018}
\bibinfo{author}{\bibfnamefont{S.}~\bibnamefont{Das}},
  \bibinfo{author}{\bibfnamefont{J.}~\bibnamefont{Riest}},
  \bibinfo{author}{\bibfnamefont{R.~G.} \bibnamefont{Winkler}},
  \bibinfo{author}{\bibfnamefont{G.}~\bibnamefont{Gompper}},
  \bibinfo{author}{\bibfnamefont{J.~K.~G.} \bibnamefont{Dhont}},
  \bibnamefont{and}
  \bibinfo{author}{\bibfnamefont{G.}~\bibnamefont{N\"{a}gele}},
  \bibinfo{journal}{Soft Matter} \textbf{\bibinfo{volume}{14}},
  \bibinfo{pages}{92} (\bibinfo{year}{2018}).

\bibitem[{\citenamefont{Jover et~al.}(2012)\citenamefont{Jover, Haslam,
  Galindo, Jackson, and M\"{u}ller}}]{Jover2012}
\bibinfo{author}{\bibfnamefont{J.}~\bibnamefont{Jover}},
  \bibinfo{author}{\bibfnamefont{A.~J.} \bibnamefont{Haslam}},
  \bibinfo{author}{\bibfnamefont{A.}~\bibnamefont{Galindo}},
  \bibinfo{author}{\bibfnamefont{G.}~\bibnamefont{Jackson}}, \bibnamefont{and}
  \bibinfo{author}{\bibfnamefont{E.~A.} \bibnamefont{M\"{u}ller}},
  \bibinfo{journal}{J. Chem. Phys.} \textbf{\bibinfo{volume}{137}},
  \bibinfo{pages}{144505} (\bibinfo{year}{2012}).

\bibitem[{\citenamefont{Hansen and Verlet}(1969)}]{Hansen1969}
\bibinfo{author}{\bibfnamefont{J.-P.} \bibnamefont{Hansen}} \bibnamefont{and}
  \bibinfo{author}{\bibfnamefont{L.}~\bibnamefont{Verlet}},
  \bibinfo{journal}{Phys. Rev.} \textbf{\bibinfo{volume}{184}},
  \bibinfo{pages}{151} (\bibinfo{year}{1969}).

\bibitem[{\citenamefont{Lin et~al.}(2003)\citenamefont{Lin, Blanco, and
  Goddard}}]{Lin2003}
\bibinfo{author}{\bibfnamefont{S.-T.} \bibnamefont{Lin}},
  \bibinfo{author}{\bibfnamefont{M.}~\bibnamefont{Blanco}}, \bibnamefont{and}
  \bibinfo{author}{\bibfnamefont{W.~A.} \bibnamefont{Goddard}},
  \bibinfo{journal}{J. Chem. Phys.} \textbf{\bibinfo{volume}{119}},
  \bibinfo{pages}{11792} (\bibinfo{year}{2003}).

\end{thebibliography}
\end{document}